\documentclass[aps,a4paper,twocolumn,english,superscriptaddress,eqsecnum,longbibliography]{revtex4-1}

\usepackage{mathtools}
\usepackage{amssymb}
\usepackage{amsmath}
\usepackage{amsfonts}
\usepackage{graphicx}
\usepackage{xcolor}
\usepackage{makecell}

\usepackage[colorlinks,
            citecolor=blue,
            linkcolor=red,
            urlcolor=blue]{hyperref}

\begin{document}

\title{Dirty bosons on the  Cayley tree:\\ Bose-Einstein condensation versus ergodicity breaking}

\author{Maxime Dupont}
\affiliation{Department of Physics, University of California, Berkeley, California 94720, USA}
\affiliation{Materials Sciences Division, Lawrence Berkeley National Laboratory, Berkeley, California 94720, USA}

\author{Nicolas Laflorencie}
\author{Gabriel Lemari\'e}
\affiliation{Laboratoire de Physique Th\'eorique, IRSAMC, Universit\'e de Toulouse, CNRS, UPS, 31062 Toulouse, France}

\begin{abstract}
    Building on large-scale quantum Monte Carlo simulations, we investigate the zero-temperature phase diagram of hard-core bosons in a random potential on site-centered Cayley trees with branching number $K=2$. In order to follow how the Bose-Einstein condensate (BEC) is affected by the disorder, we focus on both the zero-momentum density, probing the quantum coherence, and the one-body density matrix (1BDM) whose largest eigenvalue monitors the off-diagonal long-range order. We further study its associated eigenstate which brings useful information about the real-space properties of this leading eigenmode. Upon increasing randomness, we find that the system undergoes a quantum phase transition at finite disorder strength between a long-range ordered BEC state, fully ergodic at large scale, and a new disordered Bose glass regime showing conventional localization for the coherence fraction while the 1BDM displays a non-trivial algebraic vanishing BEC density together with a non-ergodic occupation in real-space. These peculiar properties can be analytically captured by a simple phenomenological description on the Cayley tree which provides a physical picture of the Bose glass regime.
\end{abstract}

\maketitle

\section{Introduction}
\label{sec:introduction}

The pioneer work of Anderson on the localization of non-interacting electrons in a random potential~\cite{anderson1958,evers2008,abrahams2010} paved the way for the study of disorder-induced phases of matter in quantum systems. Beyond a critical amount of randomness, a system can undergo drastic changes in its physical properties, generically from a \textit{delocalized} quantum state to a \textit{localized} one, such as a metal-to-insulator transition for electrons, or a superfluid-to-insulator transition for bosonic degrees of freedom \cite{dobrosavljevic2012conductor}.

In absence of interaction, the fate of electrons in a disordered environment has been, and is still intensively studied. If the transition is now well understood in finite dimension~\cite{evers2008,abrahams2010}, the case of graphs of infinite effective dimensionality, such as the Cayley tree or random graphs has recently aroused great interest~\cite{abouchacra1973, castellani1986,mirlin1994,monthus2009,biroli2012,deluca2014,monthus2011,facoetti2016,sonner2017,parisi2019,kravtsov2018,savitz2019,tikhonov2016,biroli2018,tikhonov2016b,garciamata2017,tikhonov2019,tikhonov2019b,kravtsov2015,altshuler2016,monthus2016,detomasi2019,bera2018,PhysRevB.96.201114,PhysRevB.96.064202,PhysRevE.90.052109,PhysRevB.95.104206,PhysRevLett.117.104101,pino2019, PhysRevResearch.2.012020}, due to the analogy between this problem and many-body localization (MBL) which can occur at any arbitrary energy~\cite{nandkishore2015,abanin2017,alet2018,abanin2019}.

At low-energy, the interplay of interaction and disorder in bosonic systems has received a great deal of attention following experiments on superfluid Helium in porous media~\cite{finotello1988,reppy1992} and the discovery of a novel localized phase of matter at low-temperature, the Bose glass state~\cite{halperin1986,giamarchi1987,giamarchi1988,fisher1988,fisher1989}. It can be described as an inhomogeneous gapless compressible fluid with short-ranged correlations preventing any global phase coherence responsible of delocalization properties. Known as the ``dirty boson'' problem, the localized Bose glass phase and its transition from a delocalized superfluid have been theoretically and numerically studied from one to three dimensions in various contexts~\cite{krauth1991,scalettar1991,singh1992,sorensen1992,makivic1993,weichman1995,shang1995,pai1996,herbut1997,kisker1997,rapsch1999,alet2003,prolofev2004,giamarchi2004,priyadarshee2006,hitchcock2006,weichman2007,roux2008,gurarie2009,carrasquilla2010,altman2010,lin2011,soyler2011,ristivojevic2012,iyer2012,meier2012,alvarez2013,yao2014,ristivojevic2014,alvarez2015,ng2015,doggen2017,dupont2019b}, and also reported in several experimental setups, from disordered superconductors~\cite{shahar1992,sacepe2008,sacepe2011,driessen2012,lemarie2013} to trapped ultracold atoms~\cite{white2009,deissler2010,krinner2013,derrico2014}, as well as chemically doped antiferromagnetic spin compounds~\cite{nohadana2005,hida2006,roscilde2006,hong2010,yamada2011,yu2012,yu2012b,zheludev2013,kamieniarz2016,dupont2017}.

In this paper, we investigate the low-temperature properties of strongly interacting dirty bosons on the Cayley tree. Together with an on-site random potential, the bosons have a nearest-neighbour hopping amplitude and an infinite repulsive contact interaction (hard-core constraint). This system can be efficiently simulated by extensive quantum Monte Carlo (QMC) simulations~\cite{syljuaasen2002,sandvik2010,sandvik2019}, an unbiased (``exact'') numerical method, with more than a thousand particles on the lattice for the largest system sizes accessible.

The first interest of the Cayley tree for this problem is the effective infinite dimension ($d=\infty$) of the graph, while all quantum Monte Carlo studies have focused on finite dimensional systems $d\leq 3$ so far. As discussed by Fisher {\it et al}. in their seminal work on the localization-delocalization transition for bosons~\cite{fisher1989}, it is unclear what is the correct scenario for the transition in high dimension (typically $d>4$). It is argued that there might be no finite upper critical dimension $d_\mathrm{c}$ beyond which conventional onset of mean-field theory usually takes place, and that $d_\mathrm{c}=\infty$. Contrarily, based on the exact treatment of an infinite-range hopping model~\cite{fisher1989}, which is effectively infinite dimensional, no localized phase is found, raising the question on whether or not boson localization can actually happen in high dimension. However, long-range hopping might be pathological, since the physics in the presence of disorder differs significantly from that of the short-range problem~\cite{halperin1986}. Some of these questions resonate with the problem on the Cayley tree addressed in this paper.

The second interest lies in the search of non-ergodic phases. At strong disorder, the Bose glass phase should have, as its name suggests, glassy non-ergodic properties, however they have only been little characterized (see, e.g., Refs.~\cite{PhysRevLett.91.056603, PhysRevLett.96.217203, PhysRevE.100.030102, PhysRevLett.99.050403}). The Cayley tree is one of the key models of glassy physics where the non-ergodic properties of classical disordered systems are best understood \cite{mezard1990spin,derrida1988}. Recently, the case of quantum disordered systems on the Cayley tree has attracted a strong interest. In particular, the Anderson transition on the Cayley tree presents new remarkable non-ergodic properties: The delocalized phase can be multifractal (where the states lie in an algebraically small fraction of the system) in a finite range of disorder~\cite{monthus2011,tikhonov2016,sonner2017,parisi2019, kravtsov2018,biroli2018,facoetti2016,savitz2019}, contrarily to the finite-dimensional case where multifractality appears only at criticality. Moreover the localized and critical regimes inherit a glassy non-ergodicity where the eigenstates explore only few branches~\cite{monthus2009,biroli2012, garciamata2017,PhysRevResearch.2.012020}.

Finally, we aim at comparing exact quantum Monte Carlo results to an approximate cavity mean-field approach, coming from glassy physics~\cite{mezard2001,laumann2008,krzakala2008,feigelman2010,dimitrova2011}. In particular, Feigel'man, Ioffe and M\'ezard~\cite{ioffe2010,feigelman2010} have described through this method the disorder-driven superconducting-insulator transition considering the boundaryless counterpart of the Cayley tree, the Bethe lattice. They have predicted the existence of a non-ergodic delocalized phase. Experimentally, the observation at strong disorder of large spatial fluctuations of the local order parameter in strongly disordered superconducting films \cite{sacepe2011,lemarie2013} has been interpreted as the signature of a persistence of glassy, non-ergodic properties in the superconducting phase. Although the distributions of the local order parameter observed experimentally differ from the cavity mean-field predictions on the Cayley tree \cite{lemarie2013}, these results have confirmed the importance of non-ergodic properties in this problem.

The rest of the paper is organized as follows. In Sec.~\ref{sec:model} we present the model, the numerical method, and briefly provide details about the 1BDM. In Sec.~\ref{sec:depletion}, first numerical evidences for the disorder-induced BEC depletion are presented. We then discuss microscopic aspects of the problem in Sec.~\ref{sec:real_space_properties}, where real-space properties of both off-diagonal correlations and the leading orbital are analyzed. In Sec.~\ref{sec:quantum_critical}, we look at the critical properties of the transition by performing a careful finite-size scaling, yielding estimates of the critical parameters. We then discuss the peculiar properties of the localized Bose glass regime, building on both numerical results and a phenomenological description. We finally present our conclusions and discuss some open questions in Sec.~\ref{sec:conclusion}.

\section{Model and methods}
\label{sec:model}

\subsection{Dirty hard-core bosons on Cayley trees}

We consider hard-core bosons at half-filling on a site-centered Cayley tree with $N$ lattice sites, described by the Hamiltonian
\begin{equation}
    \hat{\mathcal{H}} = -\sum\nolimits_{\langle i,j\rangle}\Bigl(\hat{b}^\dag_i\hat{b}_j+\mathrm{H.c.}\Bigr) + \sum\nolimits_{i=1}^N \mu_i \hat{n}_i,
    \label{eq:hamiltonian}
\end{equation}
where $\hat{b}^\dag_i$ ($\hat{b}_i$) is the bosonic creation (annihilation) operator on lattice site $i$, and $\hat{n}_i=\hat{b}^\dag_i\hat{b}_i$ the local density operator with the constraint $\langle\hat{n}_i\rangle\leq 1$ due to the hard-core nature of the particles, resulting from the underlying infinite repulsive interaction. The sum $\langle i,j\rangle$ restricts the tunneling to nearest-neighbor sites, and the random chemical potential $\mu_i$ is drawn from a uniform distribution $\mu_i\in[-\mu,\,+\mu]$ with $\mu$ characterizing the disorder strength. The model Eq.~\eqref{eq:hamiltonian} possesses a global continuous $\mathrm{U}(1)$ symmetry due to the conservation of its total particle number, i.e., $[\hat{\mathcal{H}},\sum_{i=1}^N\hat{n}_i]=0$.

\begin{figure}[!t]
    \centering
    \includegraphics[width=0.5\columnwidth]{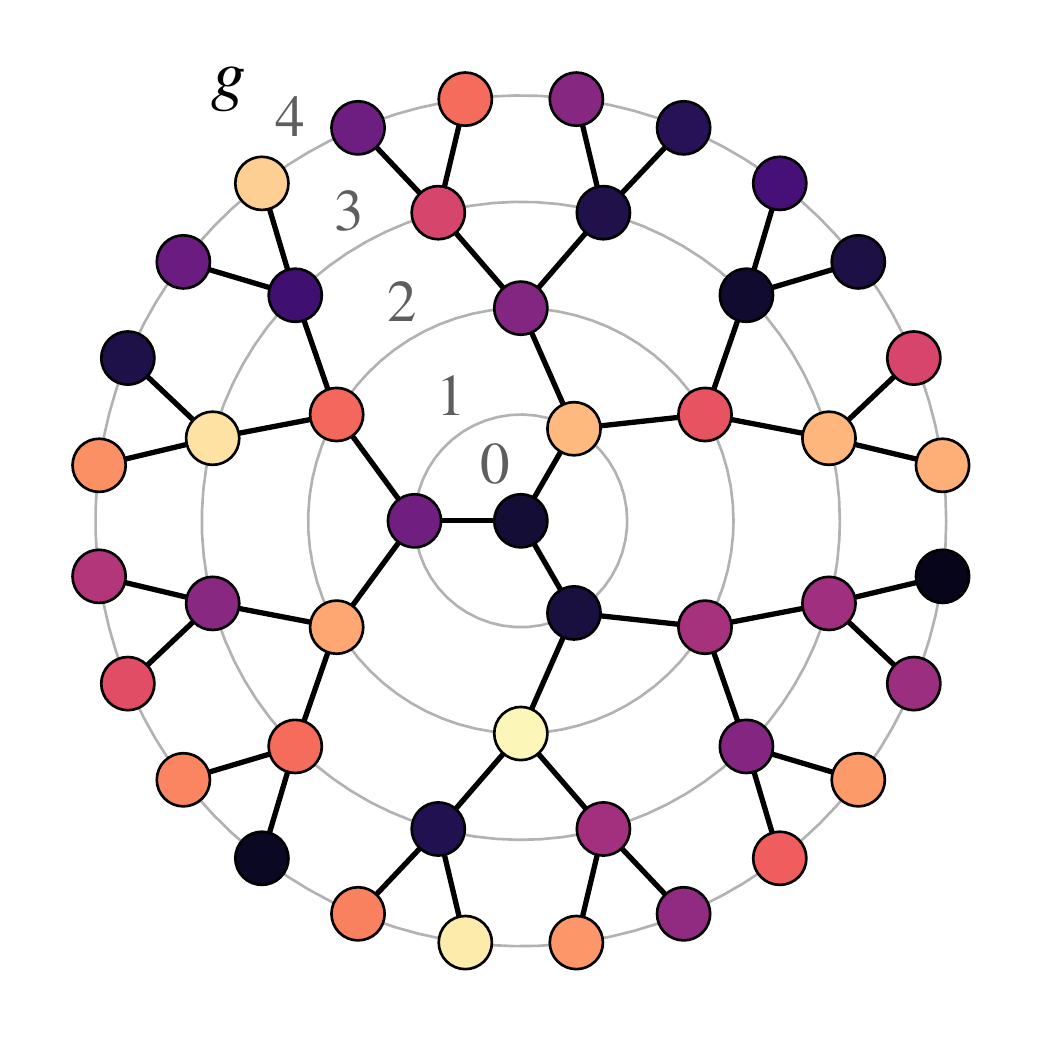}
    \caption{Site-centered Cayley tree with branching number $K=2$ and $G=4$ generations (generations from $0$ to $G=4$ are denoted by $g$). The different colors of the vertices correspond to a given random configuration of chemical potential $\mu_i$ in the Hamiltonian Eq.~\eqref{eq:hamiltonian}.}
    \label{fig:tree_nc3_ng4}
\end{figure}

The site-centered Cayley tree is defined by its branching number $K>1$ (we consider the $K=2$ case in the following) and its total number of generations $G$. See Fig.~\ref{fig:tree_nc3_ng4} for an example. The number of sites $N$ scales exponentially with $G$ as $N=1+(K+1)(K^{G}-1)/(K-1)$,
which formally mimics an infinite dimensional lattice: $N$ has the dimension of a volume and $G\sim\ln N$ of a length. Moreover, the number of lattice sites at the boundary is a finite fraction ($1-K^{-1}$ at large size) of the total number of sites, which may lead to macroscopic boundary effects to the Cayley tree, as compared its boundaryless counterpart, the Bethe lattice.

\subsection{The one-body density matrix}

\begin{figure*}[t!]
    \centering
    \includegraphics[width=2.0\columnwidth]{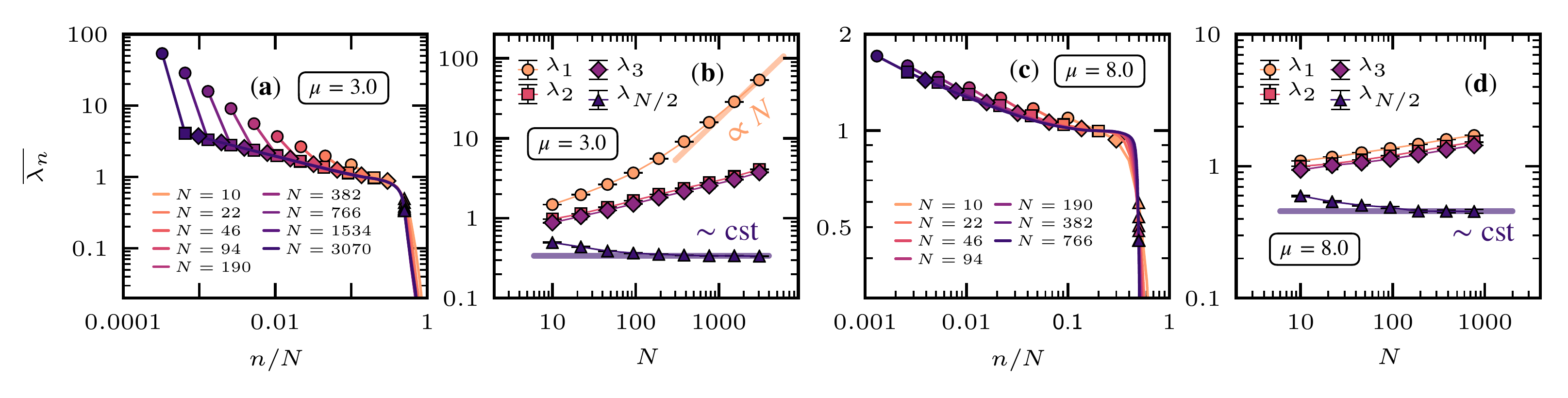}
    \caption{Disorder-averaged occupation numbers $\overline{\lambda_n}$ sorted in descending order, $\lambda_1\geq\lambda_2\geq\cdots\lambda_N$, for two disorder strengths: (a-b) $\mu=3$ and (c-d) $\mu=8$. In panels (a) and (c), the average value is plotted versus the normalized index $n/N$ for different system sizes $N$. The symbols are here to highlight specific eiegenvalues, $\overline{\lambda_1}$, $\overline{\lambda_2}$, $\overline{\lambda_3}$ and $\overline{\lambda_{N/2}}$. In panels (b) and (d), the average value is plotted versus the system size $N$ for $n=1,2,3$ and $n=N/2$. According to the Onsager-Penrose criterion~\cite{penrose1951,penrose1956,yang1962}, Bose-Einstein condensation will occur in a system which displays (at least) one occupation number of the order of $N$ as $N\to+\infty$, which is what is observed for the largest occupation number $\overline{\lambda_1}$ at $\mu=3$ in panel (b). The next occupation numbers $\overline{\lambda_2}$ and $\overline{\lambda_3}$ both have a sublinear scaling with the system size $\propto N^{0.3}$, while the middle one, $\overline{\lambda_{N/2}}$, is constant with $N$. At $\mu=8$, the first few occupation numbers, including the largest one, have a very weak sublinear scaling $\propto N^{0.1}$, and the middle one is constant. Not all occupation numbers can scale with $N$, or the constraint $\sum_{n=1}^N\lambda_n\sim O(N)$ of Eq.~\eqref{eq:loc_dens} would be violated.}
    \label{fig:occup_numbers}
\end{figure*}

A central target for this numerical study is the 1BDM, known to be an insightful object for bosonic systems, at the heart of the Penrose-Onsager criterion~\cite{penrose1951,penrose1956,yang1962} for Bose-Einstein condensation. It has also proven to be successful in the study of the high-energy many-body localization transition~\cite{bera2015,PhysRevLett.117.120402,bera2017,PhysRevB.97.104406,FIBO,hopjan2020} in one dimension. Moreover, there has been a recent proposal to measure the 1BDM for hard-core bosons in an optical lattice~\cite{pena2018}, making the quantity experimentally relevant. The 1BDM $\mathsf{C}$, defined by
\begin{equation}
    \mathsf{C}_{ij}=\Bigl\langle\hat{b}^\dag_i\hat{b}_j\Bigr\rangle,
    \label{eq:one_body_dm}
\end{equation}
is square, positive, real, and symmetric. Its diagonal elements correspond to the local densities $\langle\hat{n}_i\rangle$, such that

\begin{equation}
    \mathrm{tr}(\mathsf{C})=\sum\nolimits_{i=1}^N \langle\hat{n}_i\rangle=\bigl\langle \hat{N}_\mathrm{b}\bigr\rangle\simeq N/2,
    \label{eq:loc_dens}
\end{equation}
with $\langle\hat{N}_\mathrm{b}\rangle$ the total number of bosons in the system. The right-hand side of Eq.~\eqref{eq:loc_dens} means that we work in the grand-canonical ensemble where the particle number conservation is not enforced and therefore not restricted to half-filling, although half-filling is statistically achieved with disorder average~\cite{weichman2008}. The eigenvectors of the 1BDM Eq.~\eqref{eq:one_body_dm} are the natural orbitals,
\begin{equation}
    \mathsf{C}|\phi_n\rangle=\lambda_n|\phi_n\rangle,
    \label{eq:eigdecomp}
\end{equation}
and the eigenvalues $\lambda_n\geq 0$ are the occupation number of these orbitals with $\sum_n \lambda_n=\langle\hat{N}_\mathrm{b}\rangle$. Sorting the eigenpairs in descending order, i.e., $\lambda_1\geq\lambda_2\geq\cdots\lambda_N$, (at least) one of the eigenvalues will be of the order of the number of particles for a Bose-Einstein condensed system. This condition is known as the Onsager-Penrose criterion for Bose-Einstein condensation~\cite{penrose1951,penrose1956,yang1962}. The corresponding eigenmode $|\phi_1\rangle$ is called the leading orbital and takes the form,
\begin{equation}
    |\phi_1\rangle = \sum\nolimits_{i=1}^N a_i|i\rangle,\quad\mathrm{with}~ \sum\nolimits_{i=1}^N|a_i|^2=1,
    \label{eq:cond_mode}
\end{equation}
where $i$ designates the lattice site index. The coefficients $|a_i|^2$ account for the distribution of this leading orbital in real space.

\subsection{Numerical investigation}

\subsubsection{Quantum Monte Carlo}

So far, the few numerical studies addressing many-body interacting problems on tree-like geometries have resorted to tensor network techniques~\cite{PhysRevB.53.14004,Friedman_1997,PhysRevB.85.134415,PhysRevB.86.195137,PhysRevB.87.085107,qu2019}, but in the context of disorder-free models. Here, we instead rely on the quantum Monte Carlo method, using stochastic series expansion with directed loop updates~\cite{syljuaasen2002,sandvik2010,sandvik2019} to simulate the disordered bosonic model Eq.~\eqref{eq:hamiltonian}. For this problem, we can in practice access finite-size systems up to $G=10$ generations (coresponding to $N=3070$ lattice sites), with a sufficiently low temperature such that the algorithm is probing ground state properties. Additional informations and data are provided in App.~\ref{app:qmc_conv} regarding the convergence of our results versus the temperature. We compute the elements of the 1BDM~\cite{dorneich2001} by performing between $10^4$ and $10^5$ measurements after thermalization.

We note that the presence of ``open boundary conditions'' on the Cayley tree makes inaccessible the computation of the superfluid density~\cite{pollock1987,sandvik1997}, a very valuable quantity in the study of disorder-induced phases for bosonic systems.

\subsubsection{Disorder average}

The disorder average is performed over a large number of independent disordered samples, between $N_\mathrm{s}=300$ and $N_\mathrm{s}=2000$, depending on the system size. The exact numbers are provided in App.~\ref{app:qmc_conv}, where we also discuss the convergence of the main disorder-averaged quantities considered in this paper versus $N_\mathrm{s}$. For a physical quantity $O$, we note its disorder-averaged value $\overline{O}$ and its typical value $\exp(\overline{\ln O})$.

\section{Disorder-induced BEC depletion}
\label{sec:depletion}

\subsection{Spectrum of the one-body density matrix}

We start with an analysis of the eigenvalues of $\mathsf{C}$. In Fig.~\ref{fig:occup_numbers} the disorder-averaged occupation numbers $\overline{\lambda_n}$ with $\lambda_1\geq\lambda_2\geq\cdots\lambda_N$ is shown for various system sizes $N$ for two representative disorder strengths ($\mu=3$ and $\mu=8$)~\footnote{Because the elements of the 1BDM are computed stochastically by Monte Carlo by performing a finite number of measurements, their value is not exact, which results in negative (but close to zero) occupation numbers when diagonalizing the matrix. If calculations were to be exact, they should all be strictly positive. However, as we are interested in the first few largest occupation numbers in this paper, the presence of negative occupation numbers has little effect overall.}. At weak disorder, the first eigenvalue $\overline{\lambda_1}$ is singular, while the next ones decay smoothly to zero as the index $n$ increases. More precisely, considering the first few occupation numbers ($n=1,2,3$) and one in the middle ($n=N/2$) versus the system size, one observes that $\overline{\lambda_1}\propto N$ at large $N$, signalling Bose-Einstein condensation, according to the Onsager-Penrose criterion~\cite{penrose1951,penrose1956,yang1962}.
The next two eigenvalues $\overline{\lambda_2}$ and $\overline{\lambda_3}$ show a sublinear scaling $\propto N^{0.3}$ with the system size (this exponent decreases with the disorder strength, data not shown), and the middle one remains constant. For $\mu=8$, the largest eigenvalue has a similar behavior to $\overline{\lambda_2}$ and $\overline{\lambda_3}$, with a slow sublinear scaling $\propto N^{0.1}$ with the system size, clearly showing that no Bose-Einstein condensation occurs for this value of disorder. The middle occupation number $\overline{\lambda_{N/2}}$ is constant versus $N$. Note that because of Eq.~\eqref{eq:loc_dens}, not all eigenvalues can scale with $N$, or the constraint $\sum_{n=1}^N\lambda_n\sim O(N)$ would be violated.

\subsection{Condensed and coherent densities}

From the largest occupation number $\lambda_1$ and its relation to Bose-Einstein condensation, one can define the condensed density of bosons,
\begin{equation}
    \rho_\mathrm{cond} = \lambda_1/N.
    \label{eq:rho_cond}
\end{equation}
Having $\rho_\mathrm{cond}\sim \mathrm{constant}$ as $N\to+\infty$ is equivalent to the existence of off-diagonal long-range order in the system, associated with a spontaneous breaking of the continuous $\mathrm{U}(1)$ symmetry~\cite{yang1962}. Hence, Eq.~\eqref{eq:rho_cond} plays the role of the order parameter. In homogeneous systems with a well-defined momentum $\boldsymbol{k}$ (this is not the case of the Cayley tree), a more common (and convenient from both computational and experimental purposes) definition of the order parameter is usually based on the momentum distribution function,
\begin{equation}
    \widetilde{\rho}(\boldsymbol{k}) = \frac{1}{N^2}\sum_{\boldsymbol{r}_{ij}}e^{-i\boldsymbol{k}\cdot\boldsymbol{r}_{ij}}\Bigl\langle\hat{b}^\dag_i\hat{b}_j\Bigr\rangle,
    \label{eq:lro}
\end{equation}
with $\boldsymbol{r}_{ij}$ the distance between lattice sites $i$ and $j$. In this case, bosons generically condense into a single momentum component, the $\boldsymbol{k}=\boldsymbol{0}$ mode, meaning that $\widetilde{\rho}(\boldsymbol{0})$ serves as definition for the order parameter. In systems that are not fully translation invariant, the component with zero momentum is known as the coherent density,
\begin{equation}
    \rho_\mathrm{coh} = \frac{1}{N^2}\sum_{i=1}^N\sum_{j=1}^N\Bigl\langle\hat{b}^\dag_i\hat{b}_j\Bigr\rangle.
    \label{eq:rho_coh}
\end{equation}
However, strictly speaking, this quantity does not account for all the condensed bosons of the system Eq.~\eqref{eq:rho_cond} since what is refered as the ``glassy component'' with $\boldsymbol{k}\neq\boldsymbol{0}$ is left out~\cite{yukalov2007,astrakharchik2011,krumnow2011,muller2015}, and which results in the property that $\rho_\mathrm{coh}\leq\rho_\mathrm{cond}$. We note that the subtle links between these different quantities as well as the superfluid density (not considered in this paper) were initiated by Josephson~\cite{josephson1966}, and are still under active research, especially within the cold atom community (see Ref.~\onlinecite{muller2015} for a recent paper with references therein). Finally, although momentum is not defined on the Cayley tree, we still consider the coherent density Eq.~\eqref{eq:rho_coh} in the following, which we have found to provide relevant information on the nature of the system.

\begin{figure}[t!]
    \centering
    \includegraphics[width=1.0\columnwidth]{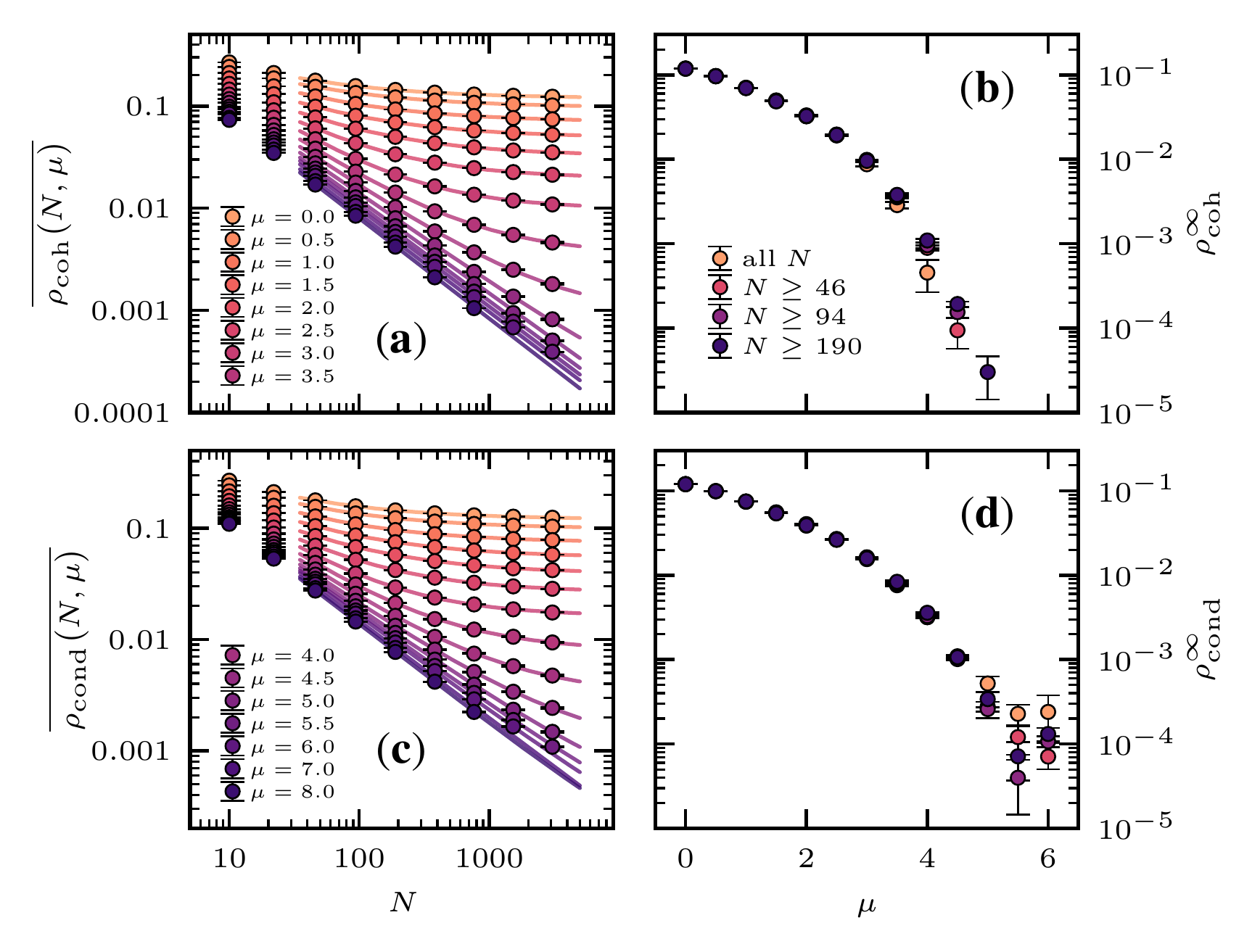}
    \caption{Left: The symbols are the disorder-averaged QMC data for (a) the coherent density of Eq.~\eqref{eq:rho_coh} and (c) the condensed density of Eq.~\eqref{eq:rho_cond}, both displayed versus the system size for various disorder strengths, as indicated on the plot. The bold lines are fits to the form Eq.~\eqref{eq:rho_fit}, taking into account all points with $N\ge 46$. These estimates $\rho^\infty(\mu)$ are shown in panels (b) and (d), for four different fitting windows. One can roughly locate a transition in the vicinity of $\mu\approx 5$ above which BEC has been destroyed.}
    \label{fig:condensate}
\end{figure}

Fig.~\ref{fig:condensate} shows the size and disorder dependence of both coherent (top) and condensate (bottom) densities. We find that, independently of the disorder strength $\mu$, they both agree with the following form
\begin{equation}
    \overline{\rho(N,\mu)} = \rho^\infty(\mu) + a(\mu)N^{-\zeta(\mu)},
    \label{eq:rho_fit}
\end{equation}
with $\rho^\infty(\mu)$, $a(\mu)$ and $\zeta(\mu)$ positive disorder-dependent parameters. The precise finite-size correction form will be discussed and analyzed in more details in Sec.~\ref{sec:quantum_critical}. Nevertheless, from this first simple analysis, one can already make important observations.
First, both coherent and condensate densities display similar behaviors, and we always observe $\overline{\rho_\mathrm{coh}}\le\overline{\rho_\mathrm{cond}}$. The extrapolated value of the coherent and condensed densities in the thermodynamic limit $\rho^\infty(\mu)$ clearly show a transition in the regime $\mu_\mathrm{c}\approx 5$. This indicates that hard-core bosons on the Cayley tree display a long-range ordered phases at small disorder, while beyond a critical disorder strength $\mu_\mathrm{c}$ the system is driven to a disordered phase where Bose-Einstein condensation has disappeared.

\subsection{Gap ratio from the largest occupation numbers}

\begin{figure}[t!]
    \centering
    \includegraphics[width=1.0\columnwidth]{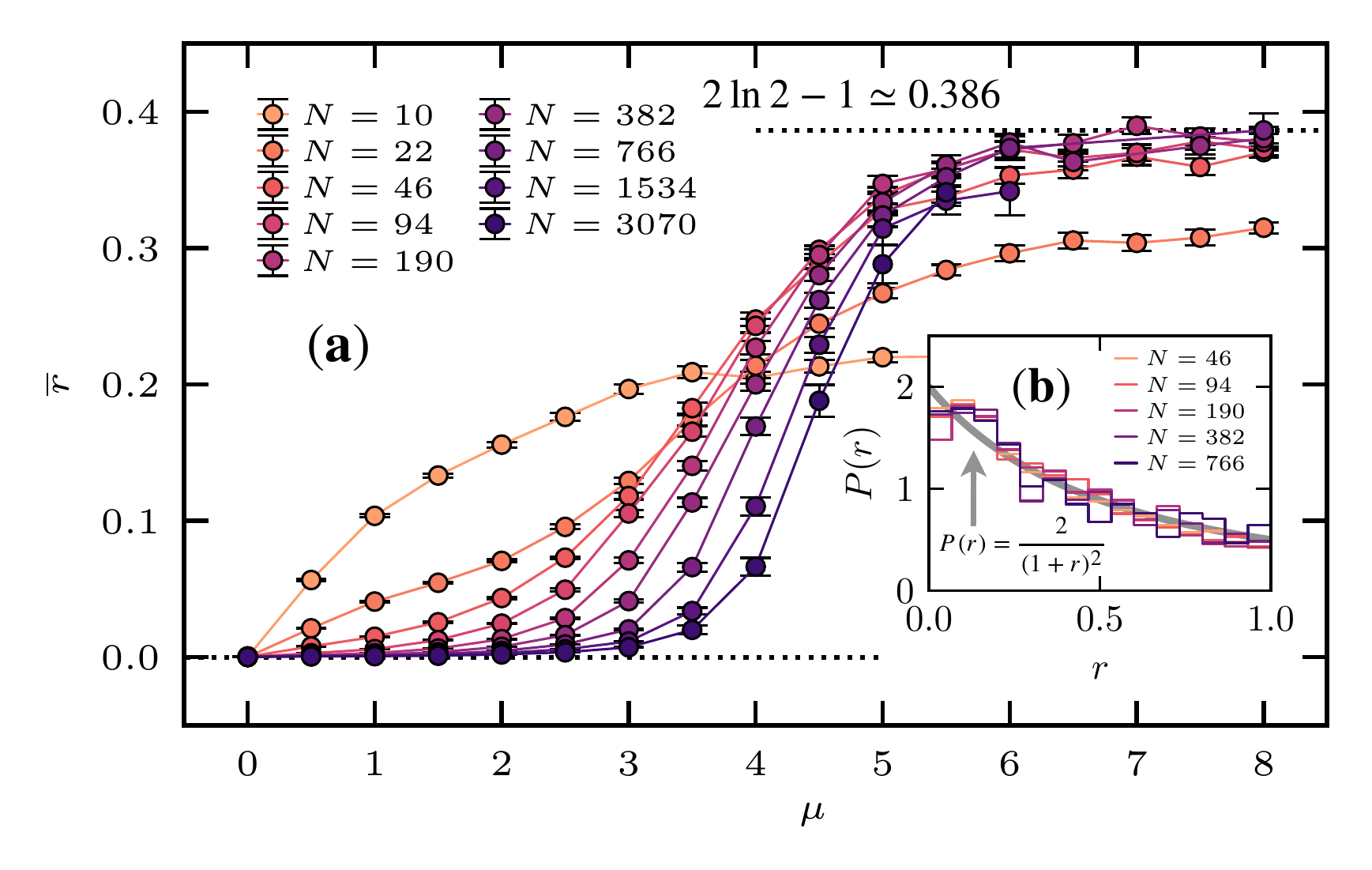}
    \caption{(a) Disorder-average adjacent gap ratio $\overline{r}$ Eq.~\eqref{eq:gap_ratio} plotted against the disorder strength $\mu$ for different system sizes $N$. The transition from the delocalized BEC phase to the disordered regime is visible around $\mu_\mathrm{c}\approx 5-6$, with clear change from $\overline{r}\to 0$ at weak disorder to the Poisson value $\overline{r}\to 2\ln 2-1\simeq 0.386$ at large disorder strength. There is a strong size dependence, except for the largest disorder strengths, where all system sizes converge onto the Poisson value.  (b) Adjacent gap ratio distribution $P(r)$ in the disordered phase, at disorder strength $\mu=8$ for system sizes $N\geq 46$. The poisson distribution $P(r)=2/(1+r)^2$, usually expected for a localized system, is in very good agreement with the numerical data.}
    \label{fig:gap_ratio}
\end{figure}

Level statistics of the eigenvalues of disordered Hamiltonians is well-knwon to be a powerful way to detect localization-delocalization transitions at high energy~\cite{PhysRevB.47.11487, PhysRevLett.79.1837,oganesyan2007}. Here, despite the fact that we work at zero temperature, one can study level statistics of the 1BDM of Eq.~\eqref{eq:one_body_dm}. More precisely, looking at the statistics of the three largest occupation numbers $\lambda_1$, $\lambda_2$ and $\lambda_3$ provides insightful information. We define the adjacent gap ratio,
\begin{equation}
    r=\frac{\mathrm{min}\bigl(\delta_1,\delta_2\bigr)}{\mathrm{max}\bigl(\delta_1,\delta_2\bigr)},
    \label{eq:gap_ratio}
\end{equation}
with $\delta_n=\lambda_{n}-\lambda_{n+1}$ the local gap between two consecutive occupation numbers. In a BEC phase, $\lambda_1\propto N$ as $N\to+\infty$, while the next occupation numbers have a sublinear scaling with $N$, as discussed in Fig.~\ref{fig:occup_numbers}. In this case, the denominator of Eq.~\eqref{eq:gap_ratio} will always scale faster with $N$ than the numerator, resulting in $\overline{r}\to 0$ in the thermodynamic limit. On the contrary, in a localized phase, one should get $r\to2\ln2-1\simeq 0.386$ if the $\lambda_n$ follow a Poisson distribution~\cite{oganesyan2007}. In Fig.~\ref{fig:gap_ratio}\,(a), these two limiting behaviors are clearly observed at small and strong disorder, respectively. In agreement with our previous analysis for the order parameters, here again one can roughly locate a transition around $\mu_\mathrm{c}\approx 5$. However, the strong size dependence of the gap ratio makes difficult a precise determination. Note that similar drifts of the gap ratio with the system size are also observed in the context of the many-body localization transition at high energy~\cite{oganesyan2007,luitz2015,bertrand2016} and the Anderson transition on random graphs \cite{biroli2012, tikhonov2016b, PhysRevResearch.2.012020}.

Here the absence of finite-size crossing signals that there is presumably no intermediate statistics at the transition, in contrast with the Anderson localization case on regular lattices~\cite{PhysRevB.47.11487, Chalker96,Bogomolny99}.
Nevertheless, Fig.~\ref{fig:gap_ratio}\,(a) confirms the existence of a spectral transition for the largest occupation numbers, from a BEC regime with $r=0$, to a disordered phase with Poisson statistics. This is also clear form the distribution $P(r)$ shown in Fig.~\ref{fig:gap_ratio}\,(b) for strong disorder ($\mu=8$), where a very good agreement is found with a Poisson distribution $P(r)=2/(1+r)^2$.

Despite the fact that spectral properties of the leading eigenvalues of the 1BDM unambiguously shows a Poisson behavior, it does not necessarily mean that the associated eigenmodes are strictly localized. Indeed, a multifractal behavior is also possible, as recently found for the MBL phase of the random-field Heisenberg chain at high energy~\cite{PhysRevLett.123.180601} (see also \cite{biroli2012, PhysRevResearch.2.012020} for the Anderson transition on random graphs). In the following section, we will address this question in a quantitive way by directly studying the localization and ergodicity properties of the leading orbital in real-space.

\section{Real-space and ergodicity properties}
\label{sec:real_space_properties}

\subsection{Local density of bosons}

\begin{figure}[!t]
    \centering
    \includegraphics[width=1.0\columnwidth]{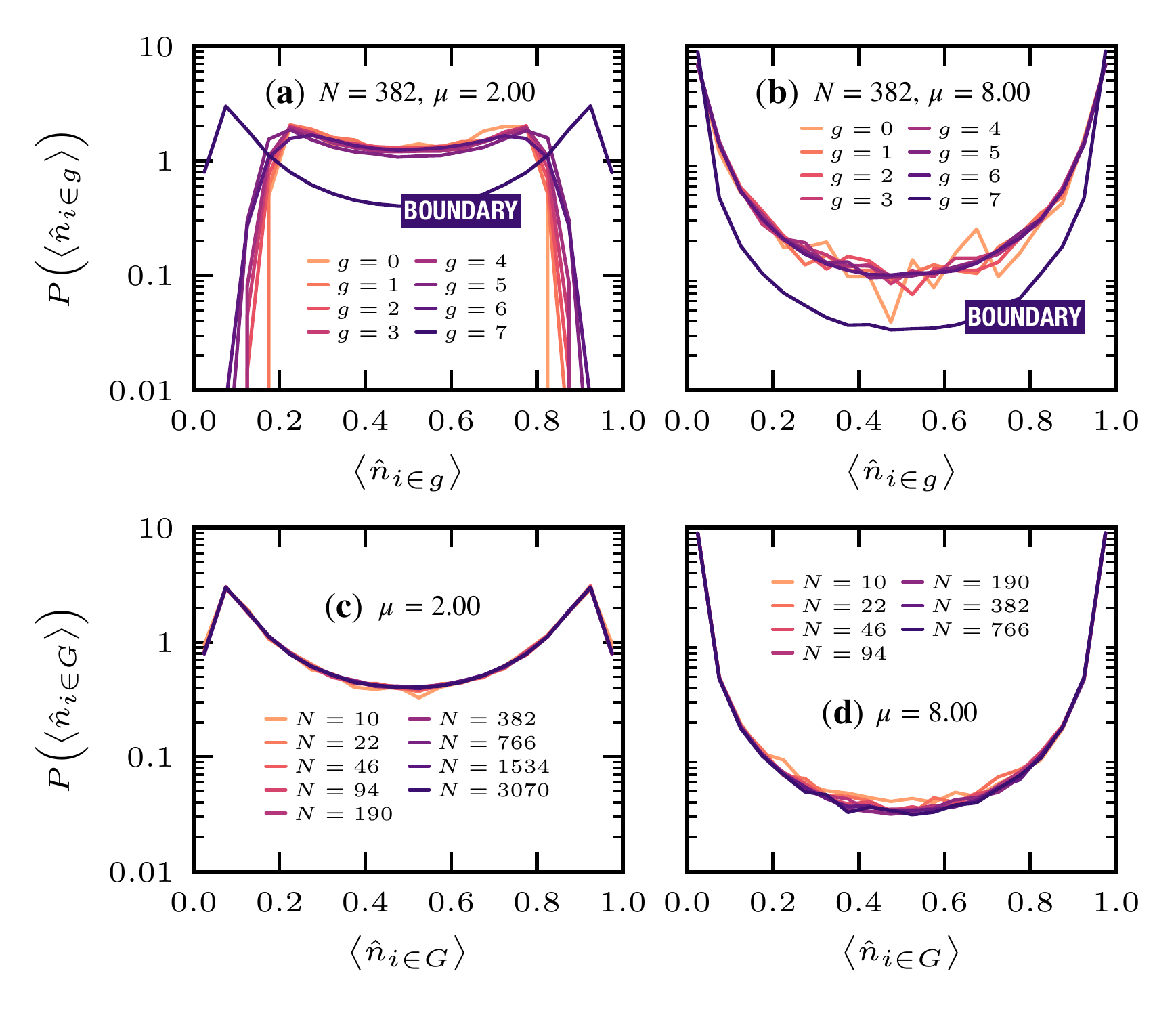}
    \caption{Probability distribution of the local density $\langle\hat{n}_i\rangle$ versus the generation $g$ to which a site $i$ belongs: $g=0$ is the center site and $g=7$ the boundary, see Fig.~\ref{fig:tree_nc3_ng4}. A system size with $G=7$ generations ($N=382$) is considered for (a) $\mu=2$ and (b) $\mu=8$. Only the densities at the boundary of the Cayley tree ($g\equiv G$) display a strong deviation from those in the bulk. They are more easily localized by having larger probabilities around the extreme values $0$ and $1$. Panels (c) and (d) show the absence of finite-size effect for the boundary sites $i\in G$.}
    \label{fig:dist_loc_density}
\end{figure}

We start this analysis by looking at the local densities $\langle\hat{n}_i\rangle$, which correspond to the diagonal entries of the 1BDM, see Eq.~\eqref{eq:loc_dens}. We show in Fig.~\ref{fig:dist_loc_density} its probability distribution for two representative disorder strengths, $\mu=2$ (Bose-Einstein condensed phase) and $\mu=8$ (disordered regime).
The sites $i$ are sorted according to the generation $g$ to which they belong. Because of the reduced connectivity of the boundary sites ($K+1$ in the bulk, and only $K-1$ at the boundary), for $g=G$ one observes a strong deviation from the occupations $\langle\hat{n}_i\rangle$ in the bulk. This is true for both phases: The probability distributions have a larger weights around the extreme values $0$ and $1$, meaning that boundary sites are more localized, as naturally expected from the reduced connectivity. However, this is not a mere finite-size effect since about half of the sites belong to the boundary on the Cayley tree with branching number $K=2$. More generally, the double-peak U-shape structure observed in Fig.~\ref{fig:dist_loc_density}\,(b) in the disordered phase is also observed in the context of many-body localization at high energy, and is a fingerprint of ergodicity breaking~\cite{lim2016,khemani2016,dupont2019,dupont2019c,laflorencie2020}.

\begin{figure}[!t]
    \centering
    \includegraphics[width=0.8\columnwidth]{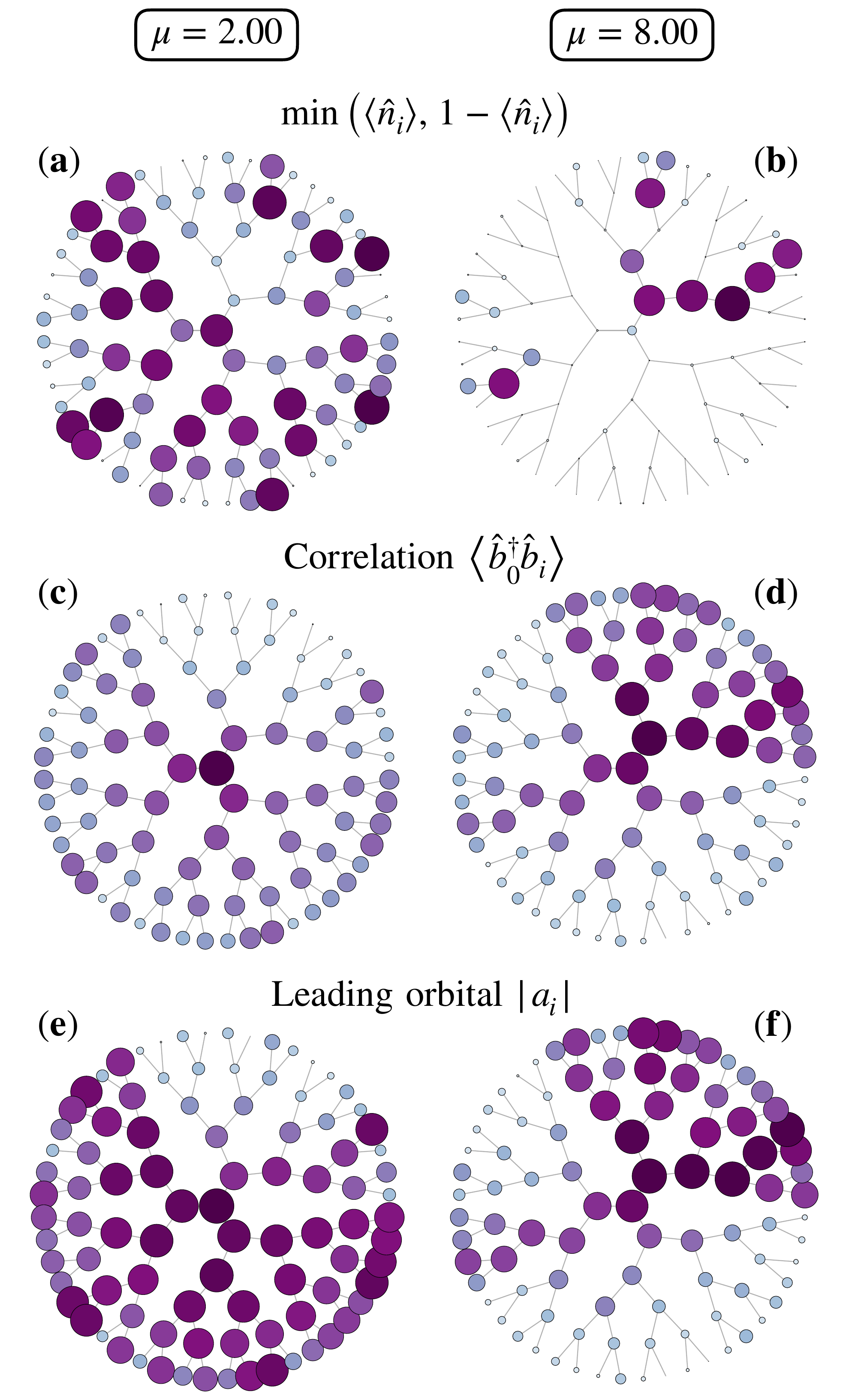}
    \caption{Real space representation of various physical quantities for a given random sample of size $N=382$ lattice sites, at small ($\mu=2$, left column) and strong ($\mu=8$, right column) disorder strengths. The scales on the panels are independent. (a-b) Deviation from perfect (non)occupation of the lattice sites measured by $\delta_i=\mathrm{min}(\langle\hat{n}_i\rangle, 1-\langle\hat{n}_i\rangle)$. The radius of the circles is proportional to $[\delta_i-\mathrm{min}(\delta_i)]/[\mathrm{max}(\delta_i)-\mathrm{min}(\delta_i)]$. (c-d) Two-point correlation $\mathsf{C}_{0i}=\langle\hat{b}^\dag_0\hat{b}_i\rangle$ from the center site, in log-scale, with the radius of the circles proportional to $[\ln \mathsf{C}_{0i}-\mathrm{min}(\ln \mathsf{C}_{0i})]/[\mathrm{max}(\ln \mathsf{C}_{0i})-\mathrm{min}(\ln \mathsf{C}_{0i})]$. (e-f) Leading orbital $|\phi_1\rangle=\sum\nolimits_{i=1}^N a_i|i\rangle$ of Eq.~\eqref{eq:cond_mode}. The radius of the circles is in log-scale and proportional to $[\ln |a_i|-\mathrm{min}(\ln |a_i|)]/[\mathrm{max}(\ln |a_i|)-\mathrm{min}(\ln |a_i|)]$. This figure is discussed throughout Sec.~\ref{sec:real_space_properties}.}
    \label{fig:sample_real_space}
\end{figure}

In Fig.~\ref{fig:sample_real_space}, we provide a real-space picture for these occupations, focusing on two representative finite-size ($G=7$, $N=382$) samples for both regimes: The BEC phase at $\mu=2$ (left column), and the disordered state at $\mu=8$ (right column). The top row displays the deviations from complete localization $\delta_i=\mathrm{min}(\langle\hat{n}_i\rangle, 1-\langle\hat{n}_i\rangle)$, from which we clearly observe that spatial inhomogeneities develop with increasing randomness. In particular, at strong disorder an apparent non-ergodic behavior settles in, with only a finite number of branches in the Cayley tree which host particle fluctuations, while a large fraction of the graph displays almost frozen sites with $\delta_i\ll 1$.

\begin{figure}[!t]
    \centering
    \includegraphics[width=1.0\columnwidth]{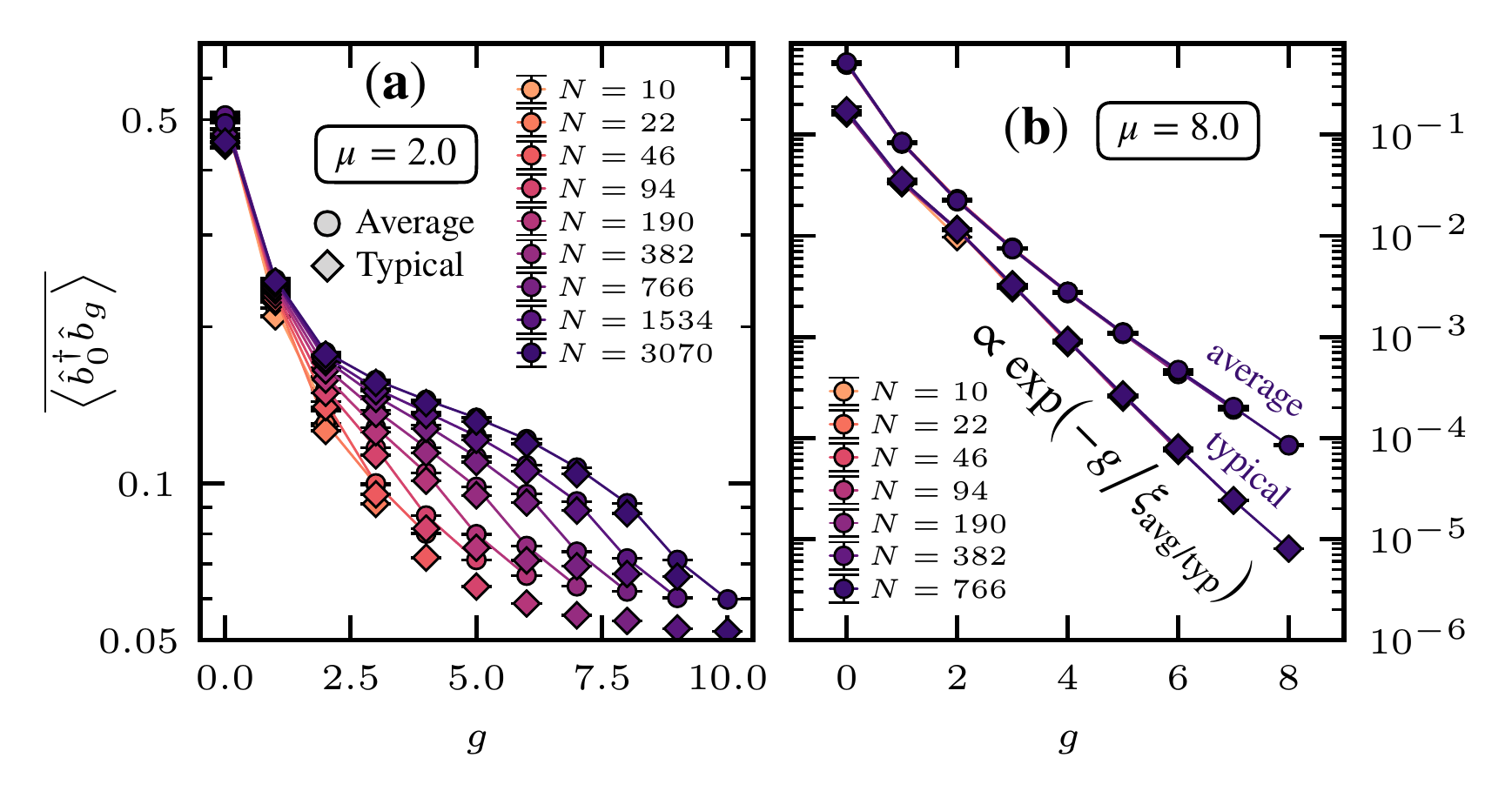}
    \caption{Disorder-averaged and typical two-point correlations between the site at the center (generation $g=0$) and the sites at the generation $g$ for different system sizes $N$ at two representative values of the disorder strength: (a) $\mu=2$ and (b) $\mu=8$. Exponential fits to the form Eq.~\eqref{eq:exp} at $\mu=8$ yield $\xi_\mathrm{avg}\approx 1.15$ and $\xi_\mathrm{typ}\approx 0.85$.}
    \label{fig:correlations}
\end{figure}

\subsection{Off-diagonal correlations}

The second row of Fig.~\ref{fig:sample_real_space} shows a snapshot of the off-diagonal correlation function measured from the root of the tree
\begin{equation}
    \mathsf{C}_{0i}\equiv\Bigl\langle\hat{b}^\dag_0\hat{b}_i\Bigr\rangle,
    \label{eq:corr}
\end{equation}
here again for two representative samples from both phases at $\mu=2$ and $\mu=8$. The spatial structure observed for the density (top row of Fig.~\ref{fig:sample_real_space}) is also clearly visible in the correlators, as shown by panels (c) and (d). Note the logarithmic scale.

\subsubsection{Average and typical correlations}

Disorder averaging has also been performed for the two-point correlation, as displayed in Fig.~\ref{fig:correlations} as a function of the distance. While the BEC phase is characterized by a slow decay at large distance towards a constant, signalling off-diagonal long-range order, the disordered regime shows short-ranged correlations with an exponential decay of the form
\begin{equation}
    \mathsf{C}_{0i}\propto\exp\bigl(-g/\xi\bigr),
    \label{eq:exp}
\end{equation}
where $g$ measures the distance between the root and site $i$. This exponential decay is clearly visible for $\mu=8$ in Fig.~\ref{fig:correlations}\;(b) where both average and typical correlators are plotted. Here two remarks are in order: (i) Finite-size effects are essentially absent in the disordered phase, in contrast with the BEC regime shown in panel (a), and (ii) while at weak disorder average and typical values are very similar (except at the boundary), in the disordered phase they decay with two different characteristic lengths $\xi_\mathrm{avg/typ}$. Such a difference between average and typical correlations is a qualitative sign of non-ergodicity (see e.g. \cite{aizenman2011absence, morone2014, tikhonov2019b, PhysRevResearch.2.012020}).

\subsubsection{Distributions}

\begin{figure}[!ht]
    \centering
    \includegraphics[width=1.0\columnwidth]{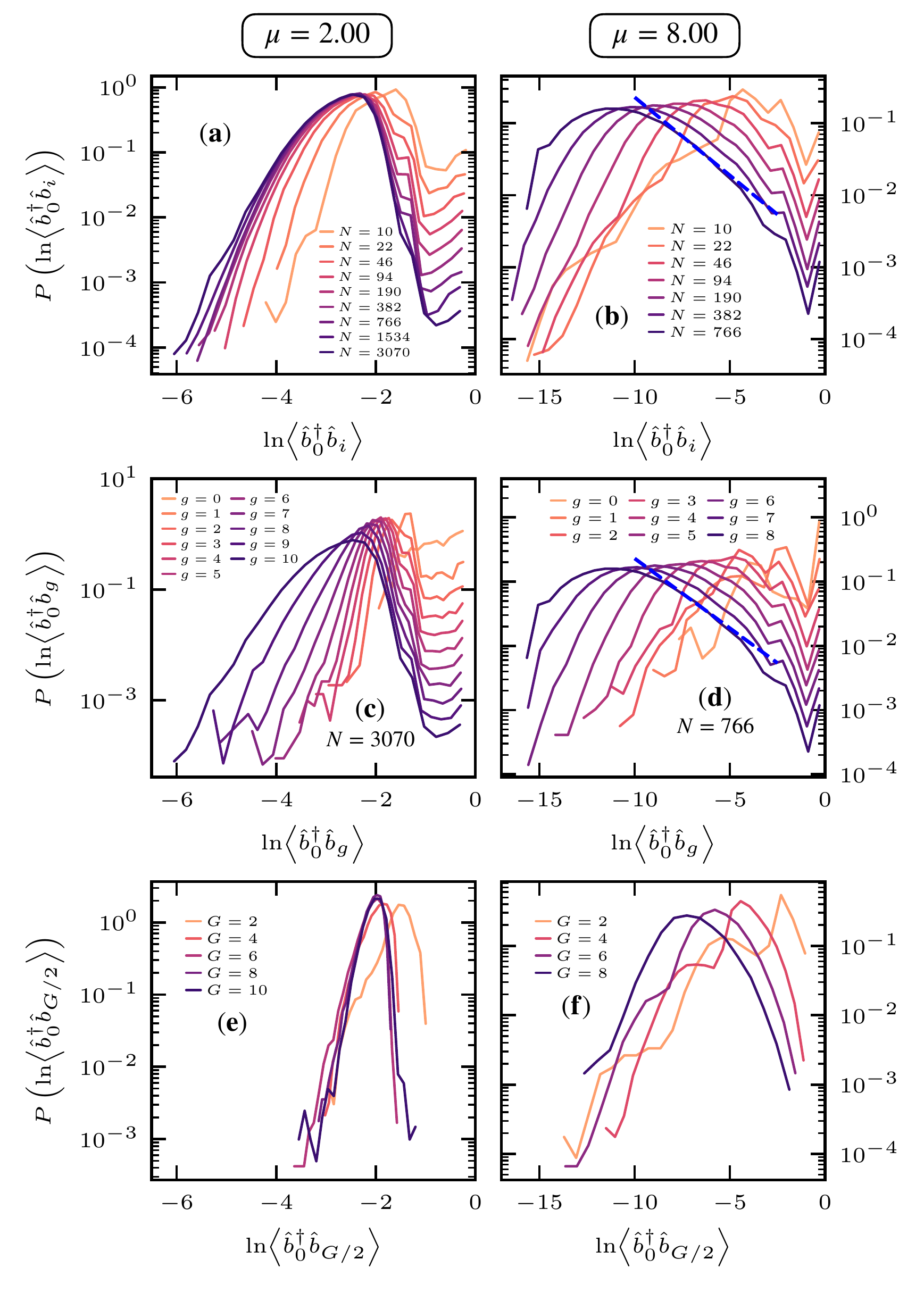}
    \caption{Distribution of different types of correlators in the Bose-Einstein condensed phase (left panels, $\mu=2$) and in the disordered phase (right panels, $\mu=8$): The correlator $\mathsf{C}_{0i}$ over all sites $i$ for different system sizes (upper panels (a) and (b)), $\mathsf{C}_{0g}$ over all sites at generation $g$ for a fixed large system size (panels (c) and (d)) and $\mathsf{C}_{0G/2}$ for different values of the total number of generations $G$ (panels (e) and (f)). In the BEC phase (left panels), a clear localizing effect of the boundary is observed in panel (c) with a sharp broadening of the distribution for $g$ close to $G=10$. In the bulk of the tree shown in panel (e), the distribution is stationary at sufficiently large system size, indicating long-range order. In the disordered phase, the distribution follows a traveling wave regime, i.e. drifts towards lower values of $\ln\mathsf{C}$ at constant speed $1/\xi_\mathrm{typ}$ with a fixed shape and a right tail close to power-law $P(\mathsf{C})\sim \mathsf{C}^{-(1+B)}$ with $B\approx 0.5$ shown by the blue dashed lines in panels (b) and (d) (see text). Such a behavior is characteristic of a non-ergodic phase~\cite{derrida1988,monthus2009,Monthus_2012,feigelman2010}.}
    \label{fig:dist_correlations}
\end{figure}

In order to better explore microscopic properties and the spatial features of the off-diagonal correlations, we show different types of distributions in Fig.~\ref{fig:dist_correlations}, again for weak ($\mu=2$) and strong ($\mu=8$) disorder. This quantity is indeed central in studies of non-ergodicity on this type of graphs~\cite{feigelman2010,morone2014,tikhonov2019, PhysRevB.101.014203,PhysRevResearch.2.012020}. We have considered the distribution of the correlator $\mathsf{C}_{0i}$ over all sites $i$ for different system sizes [panels (a) and (b) of Fig.~\ref{fig:dist_correlations}], $\mathsf{C}_{0g}$ over all sites at generation $g$ for a fixed large system size [panels (c) and (d)] and $\mathsf{C}_{0G/2}$ for different values of the total number of generations $G$ [panels (e) and (f)].

In the Bose-Einstein condensed phase at weak disorder $\mu=2$, the different correlators allow to clearly identify the localizing effect of the boundary (also seen in Fig.~\ref{fig:dist_loc_density}). Similarly to Fig.~\ref{fig:correlations} for the disorder averaged correlator $\overline{\mathsf{C}_{0g}}$ which decreases much faster close to the boundary than in the bulk of the tree, one observes in panel (c) a sharp broadening of the distribution of $\mathsf{C}_{0g}$ close to the boundary. A similar localizing effect of the boundary arises also in the Anderson localization problem on the Cayley tree~\cite{sonner2017}. On the contrary, in panel (e), the distribution of $\mathsf{C}_{0G/2}$ in the bulk of the tree reaches a stationary distribution, characteristic of long-range order. In panel (a), the correlator $\mathsf{C}_{0i}$ over all sites $i$ is clearly dominated by the boundary sites, which represent half of the total number of sites.

In the disordered phase at $\mu=8$, one clearly observes in the panels (b) and (d) a traveling wave regime where the distribution of the correlator drifts towards lower value of $\mathsf{C}_{0g}$ by always keeping the same shape at a constant speed $1/\xi_\mathrm{typ}$, where $\overline{\ln \mathsf{C}_{0g}} = -g/\xi_\mathrm{typ}$. Moreover, the distribution $P(\ln\mathsf{C})$ develops at large $g$ or $N$ a right tail close to exponential decrease $P(\ln\mathsf{C}) \sim \exp(-B\ln \mathsf{C})$ which translates into a power-law tail for $P(\mathsf{C})\sim \mathsf{C}^{-(1+B)}$ with an exponent $B\approx 0.5$. This large right tail is responsible of the different decay of the averaged and typical correlator, see Fig.~\ref{fig:correlations}\;(b). Such a behavior is characteristic of a non-ergodic phase and is often related to the characteristic directed polymer physics on the Cayley tree \cite{derrida1988,monthus2009,Monthus_2012,feigelman2010}. In this context, an exponent $B<1$ signals replica symmetry breaking, a characteristic glassy property \cite{mezard1990spin}. The disordered regime can therefore be seen as a Bose glass.

\subsection{Ergodicity properties of the leading orbital}

The leading orbital $|\phi_1\rangle=\sum\nolimits_{i=1}^N a_i|i\rangle$, associated to the largest eigenvalue of the 1BDM, is the most delocalized one, corresponding to the condensed mode in the BEC regime. In the last row of Fig.~\ref{fig:sample_real_space}, we represent the weights $|a_i|$ in real space for the same samples as in the above rows of the same figure with $\mu=2$ (left) and $\mu=8$ (right). It is quite remarkable that the same spatial structure observed for the correlators in the middle panels also emerges for this leading orbital.

In order to be more quantitative, we study the participation entropy $S_q$~\cite{luitz2014}, derived from the $q$th moments of the eigenmode $|\phi_1\rangle$. This quantity informs us on its (de)localization properties in real space. It is defined by,
\begin{equation}
    S_q = \frac{1}{1-q}\ln\Biggl(\sum\nolimits_{i=1}^N|a_i|^{2q}\Biggr).
    \label{eq:part_ent_sq}
\end{equation}
In the thermodynamic limit, one gets $S_q=\ln N$ for a perfectly delocalized mode whereas $S_q=\mathrm{constant}$ if the mode is localized. In an intermediate situation, $S_q\propto\ln (N^{D_q})\propto D_q\ln N$ with $0<D_q<1$ called the (possibly $q$-dependent) (multi)fractal dimension. In this case, the mode is delocalized (the participation entropy still grows with $N$) but non-ergodic (the scaling is slower than in the perfectly delocalized case, meaning that it does not occupy uniformly the whole space). The extreme cases $D_q=1$ and $D_q=0$ correspond to perfect delocalization and localization, respectively.

\begin{figure}[t!]
    \centering
    \includegraphics[width=1.0\columnwidth]{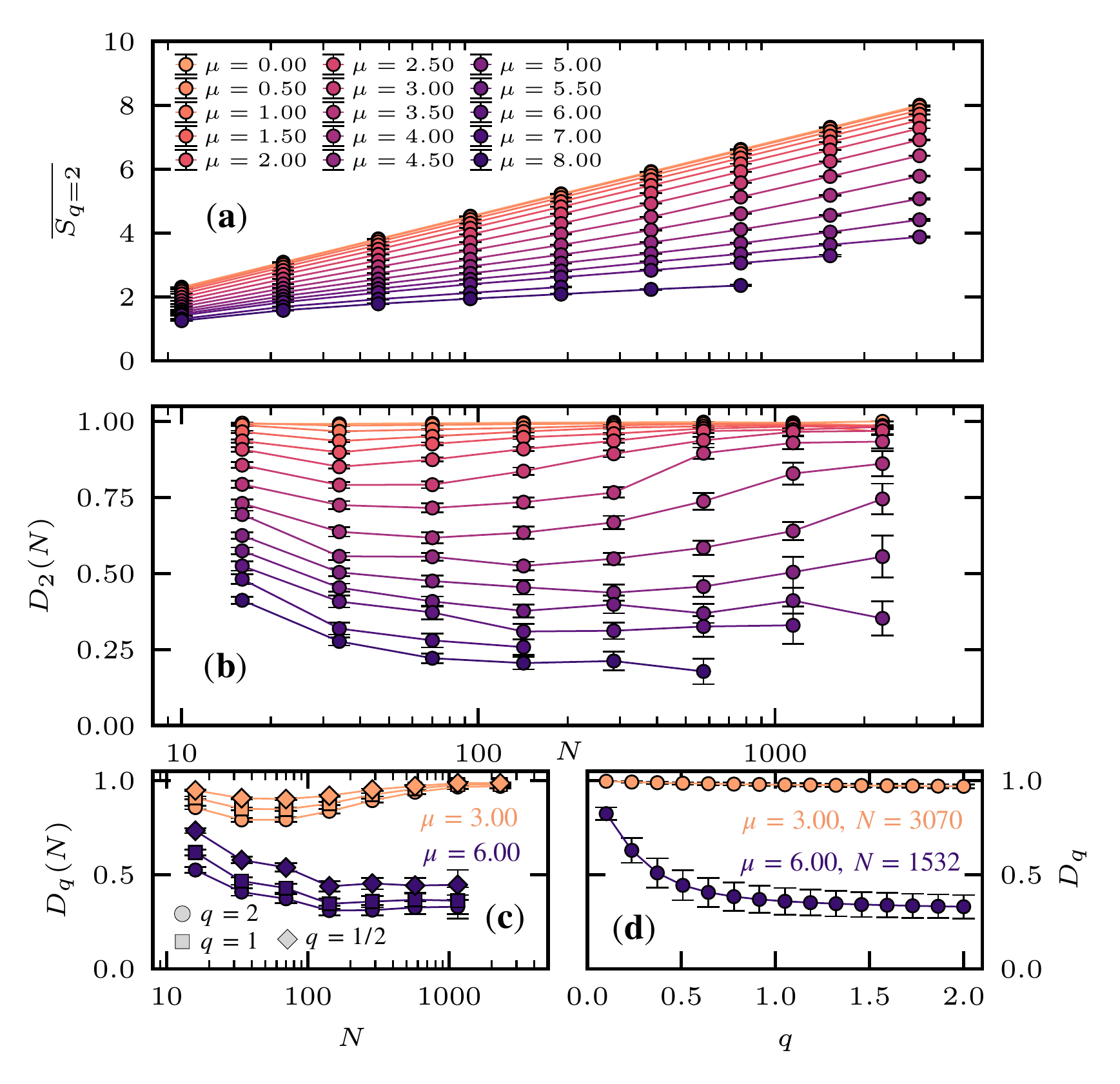}
    \caption{(a) Disorer-averaged participation entropy $\overline{S_{q=2}}$ of the leading orbital versus the system size $N$ for various disorder strengths $\mu$, see definition of Eq.~\eqref{eq:part_ent_sq}. (b) Local slope of the disorer-averaged participation entropy Eq.~\eqref{eq:mult_frac_dim} versus the system size is displayed, and should saturate to $D_2(\mu)$ in the limit $N\to+\infty$. At small disorder, we observe a non-monotonous behavior characterized by a minimum before getting $D_2\simeq 1$ as the system size is increased. At strongder disorder, $D_2$ seems to saturate to a finite value smaller than one, signaling nonergodicity. (c) Local slope of the disorer-averaged participation entropy of index $q$ ($q=0.5$, $q=1$ and $q=2$) versus the system size $N$ for $\mu=3$ and $\mu=6$. No $q$ dependence is observed at small disorder as $N\to+\infty$, while $D_q(N)$ saturates to slightly different values depending on $q$, suggesting multifractality. (d) Same data as panel (b) for a fixed system size $N$ at $\mu=3$ and $\mu=6$, versus the index $q$. Multifractality is confirmed at strong disorder, with $D_q$ being $q$-dependent.}
    \label{fig:ent_cond_mode}
\end{figure}
In the following, we mainly focus on the $q=2$ case, which recovers the usual inverse participation ratio (IPR) with $S_2=-\ln(\mathrm{IPR})$~\cite{visscher1972}. We show in Fig.~\ref{fig:ent_cond_mode}\,(a) the disorder-averaged participation entropy $\overline{S_2}$ of the leading mode versus the system size $N$ for various disorder strengths $\mu$. As expected, we observe a logarithmic increase $\ln N$, with a prefactor which seems to gradually change with increasing randomness. While the multifractal dimension is defined in the thermodynamic limit, it is instructive to consider its finite-size version at fixed $N$ to then try to extract its $N\to +\infty$ value,
\begin{equation}
    D_q\bigl(N\bigr) = \frac{\mathrm{d}\overline{S_q\bigl(N\bigr)}}{\mathrm{d}\ln N}\quad\mathrm{with}\quad D_q\equiv D_q\bigl(N\to\infty\bigr).
    \label{eq:mult_frac_dim}
\end{equation}
This local slope is displayed in Fig.~\ref{fig:ent_cond_mode}\,(b) as a function of system size $N$. In absence of disorder, the leading orbital is perfectly delocalized with $D_2(N)=1$ for all system sizes. When introducing disorder in the system, the local slope becomes non-monotonous by developing a minimum at $N=N^*(\mu)$ before increasing towards $D_2(N)\to 1$ with system size. Such a behavior was also observed in the IPR on random regular graphs for the Anderson localization transition~\cite{tikhonov2016b, garciamata2017, biroli2018}. This gives rise to an additive correction to the scaling of the participation entropy at large $N$, $S_q=\ln N+b_q$, with $b_q<0$. This negative constant correction can be physically related to a finite nonergodicity volume $\Lambda_{S_q}=\exp(-b_q)$, as argued later. The position $N^*(\mu)$ of the minimum increases with the disorder, resulting in the nonergodicity volume also increasing with $\mu$. A detailed scaling analysis will be performed below, in Sec.~\ref{sec:quantum_critical}.

\begin{figure*}[t]
    \centering
    \includegraphics[width=\linewidth]{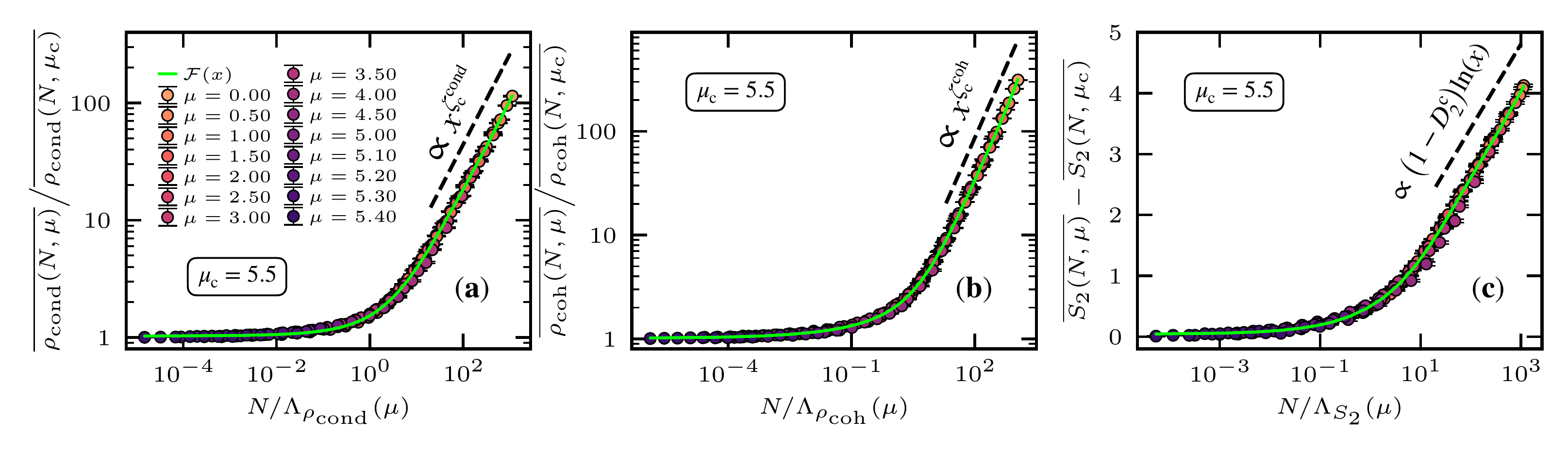}
     \caption{Finite-size scaling analysis of the disorder-averaged (a) condensed density $\overline{\rho_\mathrm{cond}}$, (b) coherent density $\overline{\rho_\mathrm{coh}}$, and (c) participation entropy of the leading orbital $\overline{S_2}$ in the regime $\mu\leq\mu_\mathrm{c}$. The best scaling of the data is obtained for a volumic scaling Eq.~\eqref{eq:scaling_linvol} at $\mu_\mathrm{c}\approx 5.5$. The green curve shows the scaling function $\mathcal{F}(N/\Lambda)$ of Eq.~\eqref{eq:scafunc} fitted to the data, with a quantity-dependent and disorder-dependent scaling parameter $\Lambda(\mu)$. The divergence of the non-ergodicity volume $\Lambda$ at criticality is shown in Fig.~\ref{fig:collapse_lbda}. The dashed lines correspond to the behavior of the scaling function for the three quantities for $N\gg\Lambda$, according to Eq.~\eqref{eq:Frho_large_x} and Eq.~\eqref{eq:FS2_large_x}.}
    \label{fig:collapse}
\end{figure*}

At stronger disorder, in the regime where the BEC order parameter was found to vanish, we clearly observe a different behavior for the prefactor $D_2$, with an apparent saturation at a value $D_2<1$, thus signalling that the leading orbital associated to the most delocalized mode is no longer ergodic on the Cayley tree, but rather (multi)fractal. Panel (c) of Fig.~\ref{fig:ent_cond_mode} shows such a difference between the two regimes, for three values of the R\'enyi parameter $q=0.5,\, 1,\, 2$. For $\mu=3$ (BEC phase) full ergodicty of the leading orbital is recovered for large enough system size with $D_q\to 1$. Instead, in the disordered regime at $\mu=6$, $D_q$ clearly saturates to a non-ergodic value, with an additional signs of multifracatality as a non-trivial $q$-dependence is found. This is better visible in Fig.~\ref{fig:ent_cond_mode} where one also sees strong multifractality at small $q$ (as also observed for the Anderson transition in infinite dimension~\cite{sonner2017, PhysRevResearch.2.012020}, or for gapped ground-states of spin chains~\cite{PhysRevLett.112.057203}), followed by an almost $q$-independent regime.

\section{Quantum critical properties}
\label{sec:quantum_critical}

\subsection{Scaling analysis across the transition}

We established in the previous sections that a transition takes place in the system around $\mu_\mathrm{c}\approx 5-6$ between a Bose-Einstein condensate at small disorder and a disordered phase at stronger disorder. We now turn our attention to a finite-size scaling analysis of the various quantities across the transition in order to characterize it. We will focus on the long-range ordered phase side of the phase diagram with $\mu\leq\mu_\mathrm{c}$. In the disordered phase at $\mu>\mu_\mathrm{c}$, the numerical simulations are limited in size and strength of the disorder so that we could not perform a conclusive finite-size scaling analysis.

The finite-size scaling analysis of localization transitions in graphs of effective infinite dimensionality such as the Cayley tree is particularly subtle. This was illustrated recently in the Anderson transition on random graphs \cite{garciamata2017, tikhonov2019, PhysRevResearch.2.012020}, in the MBL transition \cite{PhysRevLett.123.180601, laflorencie2020} and in certain classes of random matrices \cite{pino2019,kravtsov2020,khaymovich2020}. The difficulty comes from the fact that the volume of the system $N$ (the number of sites) varies exponentially with the linear size of the system, i.e., the number of generations $G$ in the Cayley tree. This implies that a volumic scaling law $\mathcal{F}(N/\Lambda)$ depending on the ratio of volume $N$ by a characteristic volume $\Lambda$ (e.g., a correlation volume) is distinct from a linear scaling $\mathcal{F}(G/\xi)$ depending on the ratio of the size $G$ over the characteristic length $\xi$. These different types of scaling have important implications for the nature of the transition and of the different phases. In particular, a linear scaling can imply (depending on the critical behavior) a non-ergodic delocalized phase, see Refs.~\cite{garciamata2017, PhysRevResearch.2.012020, PhysRevLett.123.180601,laflorencie2020}.

We carried out a detailed scaling analysis of the behavior of $\overline{S_2}$, $\overline{\rho_\mathrm{cond}}$ and $\overline{\rho_\mathrm{coh}}$ according to the size of the system, and tested these various scaling assumptions (linear and volumic). The results show a quantitative agreement of the data with a volumic scaling assumption, attesting to the ergodic character of the delocalized phase, with compatible values of $\mu_\mathrm{c}$ and critical exponent for $\overline{S_2}$, $\overline{\rho_\mathrm{cond}}$ and $\overline{\rho_\mathrm{coh}}$. We detail this analysis in this section. The approach we have used is very similar to what has been done in the context of Anderson localization on random graphs and the MBL transition \cite{garciamata2017, PhysRevResearch.2.012020, PhysRevLett.123.180601,laflorencie2020}. We assume some value of $\mu_\mathrm{c}$ which belongs to the set of $\mu$'s that we have simulated. We then consider the scaling observable $\mathcal O \equiv \overline{\rho(N,\mu)}/\overline{\rho(N,\mu_\mathrm{c})}$ for $\rho_\mathrm{cond}$ and $\rho_\mathrm{coh}$ and $\mathcal O \equiv \overline{S_2\bigl(N,\mu\bigr)}-\overline{S_2\bigl(N,\mu_\mathrm{c}\bigr)}$ for $S_2$ (the substraction instead of the division by the critical behavior comes from the fact that $S_2$ is an entropy) and test the validity of the volumic or linear scaling assumptions:
\begin{equation}
    \mathcal{O} = \mathcal{F}^\mathrm{vol}_\mathcal{O}\Bigl(N/\Lambda\Bigr)\quad\mathrm{or}\quad\mathcal{O} = \mathcal{F}^\mathrm{lin}_\mathcal{O}\Bigl(G/\xi\Bigr).
    \label{eq:scaling_linvol}
\end{equation}
To do this, we perform a Taylor expansion of the scaling functions around $\mu\equiv\mu_\mathrm{c}$~{\cite{PhysRevLett.82.382, PhysRevB.84.134209}}:
\begin{equation}\label{eq:scafunc}
    \mathcal{F}\Bigl(\Theta\mathcal{N}^{1/\nu}\Bigr) = \sum_{j=0}^n a_j\Bigl(\Theta\mathcal{N}^{1/\nu}\Bigr)^j,
\end{equation}
with $\mathcal{N}=N$ the volume or $\mathcal{N}=G$ the depth of the tree and
\begin{equation}
    \Theta =\bigl(\mu-\mu_\mathrm{c}\bigr)+\sum_{j=2}^m b_j\bigl(\mu-\mu_\mathrm{c}\bigr)^j\;,
\end{equation}
The orders of expansion have been set to $n=5$ and $m=3$. Therefore, $N_\mathrm{dof}=n+m+1$ parameters are to be fitted (including the critical exponent $\nu$). The goodness of fit (calculated from the chi-squared statistic divided by the number of degrees of freedom) should be of order one for an acceptable fit.

A systematic test of different choices of $\mu_\mathrm{c}$ and volumic and linear scaling hypotheses is represented in Fig.~\ref{fig:chi2_collapse} of the appendix. It gives a clear indication that the data in the condensed phase $\mu<\mu_\mathrm{c}$ are compatible with a volumic scaling with $\mu_\mathrm{c}\approx 5.5(5)$. The critical behavior at $\mu=5.5\approx \mu_\mathrm{c}$ is well described by
\begin{equation}
    \overline{\rho\bigl(N,\mu_\mathrm{c}\bigr)}\sim N^{-\zeta_c},
\end{equation}
and
\begin{equation}
    \overline{S_2\bigl(N,\mu_\mathrm{c}\bigr)}\sim D_2^\mathrm{c} \ln N + b_2^\mathrm{c},
\end{equation}
with $\zeta_\mathrm{c}^\mathrm{cond}\approx 0.80(5)$, $\zeta_\mathrm{c}^\mathrm{coh}\approx 0.95(3)$, and a fractal dimension $D_2^\mathrm{c}\approx 0.38(6)$. The error bars are estimated by considering different $\mu_\mathrm{c}$ within the range $\mu_\mathrm{c}=5$ and $\mu_\mathrm{c}=6$. The volumic scaling behavior, represented in Fig.~\ref{fig:collapse}, together with the ergodic behavior at small $\mu$, $\overline{\rho_\mathrm{cond}}\approx\rho^\infty_\mathrm{cond}$ and $\overline{\rho_\mathrm{coh}}\approx\rho^\infty_\mathrm{coh}$ and $\overline{S_2}=\ln N+b_2$ predicts an ergodic condensed phase for $N\gg \Lambda$. Indeed, the scaling function $\mathcal{F}$ behave as
\begin{equation}
    \mathcal F_\rho\bigl(x\bigr)\sim x^{\zeta_\mathrm{c}}\quad\mathrm{for}~x\gg 1,
    \label{eq:Frho_large_x}
\end{equation}
for the condensed and coherent densities, and as
\begin{equation}
    \mathcal F_{S_2}\bigl(x\bigr)\sim\bigl(1-D_2^\mathrm{c}\bigr)\ln (x)\quad\mathrm{for}~x\gg 1,
    \label{eq:FS2_large_x}
\end{equation}
for the participation entropy of the leading orbital, as shown by the dashed lines in Fig.~\ref{fig:collapse}.

\begin{figure}[!b]
    \centering
    \includegraphics[width=1.0\columnwidth]{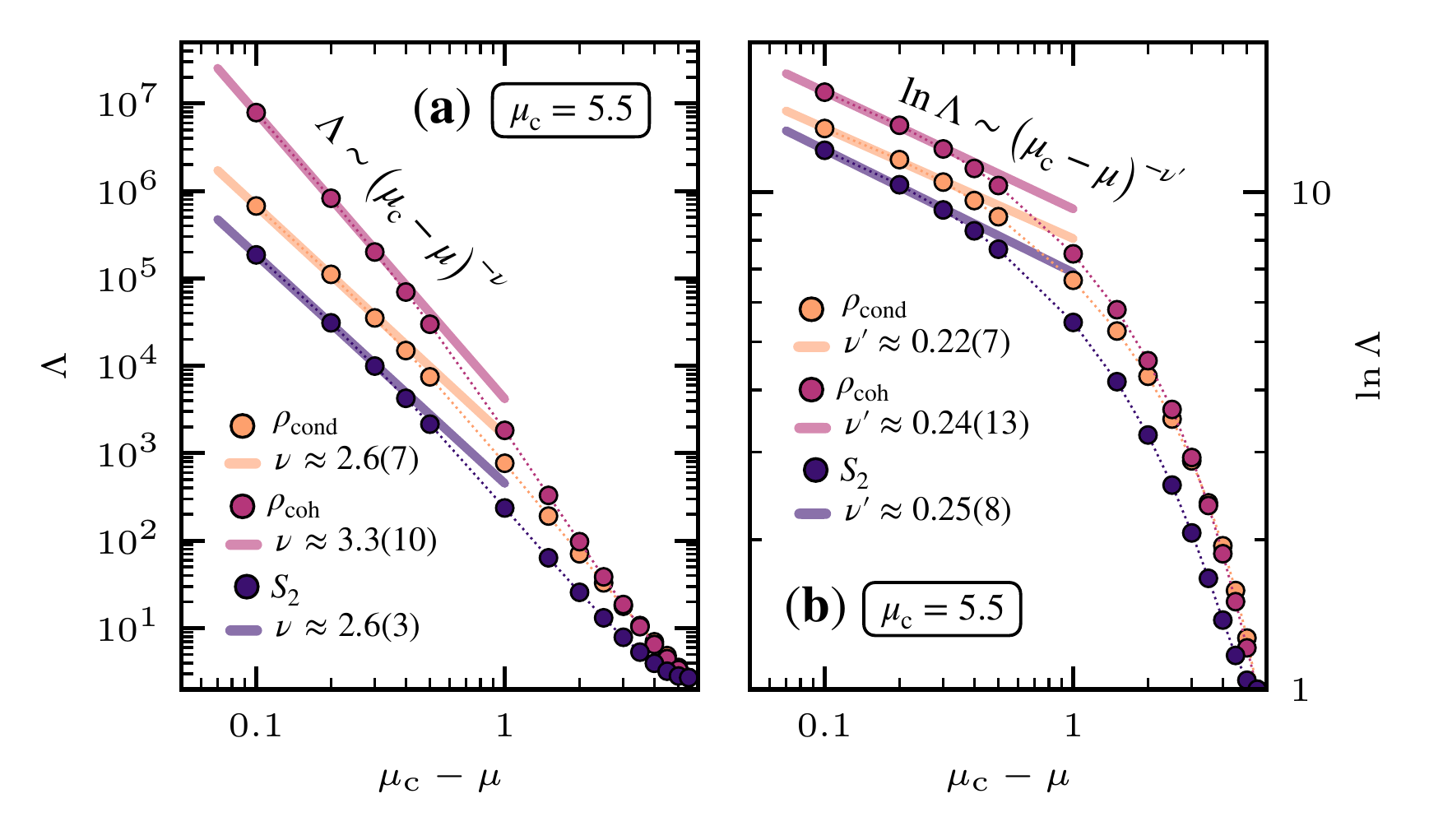}
    \caption{Divergence of the characteristic scaling volumes $\Lambda$ of the three disorder-averaged observables $\overline{\rho_\mathrm{cond}}$, $\overline{\rho_\mathrm{coh}}$ and $\overline{S_2}$ obtained from Fig.~\ref{fig:collapse}. (a) Test of an algebraic divergence of $\Lambda$: Power-law fit $\Lambda\sim (\mu_\mathrm{c}-\mu)^{-\nu}$ close to $\mu_\mathrm{c}$. The extracted critical exponents are given with the corresponding error bar, estimated from the change of $\nu$ with the choice of $\mu_\mathrm{c}$ between $\mu_\mathrm{c}=5$ and $\mu_\mathrm{c}=6$. (b) Similar to the other panel with a test of an exponential divergence of $\Lambda$, where $\ln \Lambda$ is fitted to $\ln \Lambda \sim (\mu_\mathrm{c}-\mu)^{-\nu'}$.}
    \label{fig:collapse_lbda}
\end{figure}

\subsection{Critical exponents}

\subsubsection{Correlation volumes}

The divergence of the correlation volumes $\Lambda$ is shown in Fig.~\ref{fig:collapse_lbda}. It is difficult to conclude whether the scaling volume diverges exponentially or algebraically at the transition. On the one hand, our approach presupposes an algebraic divergence, see Eq.~\eqref{eq:scafunc}, but we considered non-linear corrections so that it can describe also an exponential divergence. On the other hand, volumes vary exponentially with lengths on the Cayley tree, and if the divergence of $\Lambda$ is to be associated with an algebraic divergence of a characteristic length, then $\Lambda$ must diverge exponentially. In the panel (a) of Fig.~\ref{fig:collapse_lbda}, algebraic fits of the three $\Lambda$s as a function of $(\mu_\mathrm{c}-\mu)$ give exponents $\nu\approx 2.6$ for $\overline{\rho_\mathrm{cond}}$ and $\overline{S_2}$ and $\nu\approx 3.3$ for $\overline{\rho_\mathrm{coh}}$. These values are quite large and may suggest an exponential divergence, which is shown in panel (b) of Fig.~\ref{fig:collapse_lbda}. There, $\ln\Lambda$ versus $(\mu_\mathrm{c}-\mu)$ are fitted by a power-law with a common exponent $\nu'\approx 0.25$ for all three observables $\overline{\rho_\mathrm{cond}}$, $\overline{\rho_\mathrm{coh}}$ and $\overline{S_2}$. This scaling analysis confirms the ergodic nature of the condensed phase at small $\mu<\mu_\mathrm{c}$ when the system volume $N\gg \Lambda$, with $\Lambda$ a characteristic volume diverging at a critical value of the disorder $\mu_\mathrm{c} \approx 5.5(5)$.

\subsubsection{Order parameter}

The order parameter usually vanishes at criticality as
\begin{equation}
    \rho\sim\bigl| \mu-\mu_\mathrm{c}\bigr|^{2\beta},
\end{equation}
which defines the critical exponent $\beta$. Assuming that the characteristic volume diverges as
\begin{equation}
    \Lambda\sim\bigl|\mu-\mu_\mathrm{c}\bigr|^{-\nu},
\end{equation}
we expect
\begin{equation}
    \rho\sim \Lambda^{-2\beta/\nu}.
\end{equation}
Therefore, at criticality, for finite-size $N\leq\Lambda$, we have
\begin{equation}
    \rho\sim N^{-2\beta/\nu},
    \label{eq:critic}
\end{equation}
thus leading to the simple identification
\begin{equation}
    \zeta=2\beta/\nu.
\end{equation}
If instead of a power-law, the correlation volume diverges exponentially, see Fig.~\ref{fig:collapse_lbda}\;(b), following
\begin{equation}
    \ln\Lambda\sim\bigl|\mu-\mu_\mathrm{c}\bigr|^{-\nu'},
\end{equation}
that would imply an exponential vanishing of the order parameter
\begin{equation}
    \ln\rho\sim\bigl|\mu-\mu_\mathrm{c}\bigr|^{\beta'},
    \label{eq:expbeta}
\end{equation}
in order to verify the observed critical algebraic decay Eq.~\eqref{eq:critic}.

From our numerics, a direct estimate of the order parameter exponent is very difficult, as seen in Fig.~\ref{fig:condensate} (right panels). However, it is also clear that the vanishing of the order parameter is very fast, suggesting a large value of $\beta$, in agreement with the large value of $\nu$ (Fig.~\ref{fig:collapse_lbda}). QMC data would also be compatible with an exponential decay Eq.~\eqref{eq:expbeta}.

\subsection{Strong disorder regime}
\label{sec:loc}

\subsubsection{QMC results}

\begin{figure}[!t]
    \centering
    \includegraphics[width=\columnwidth]{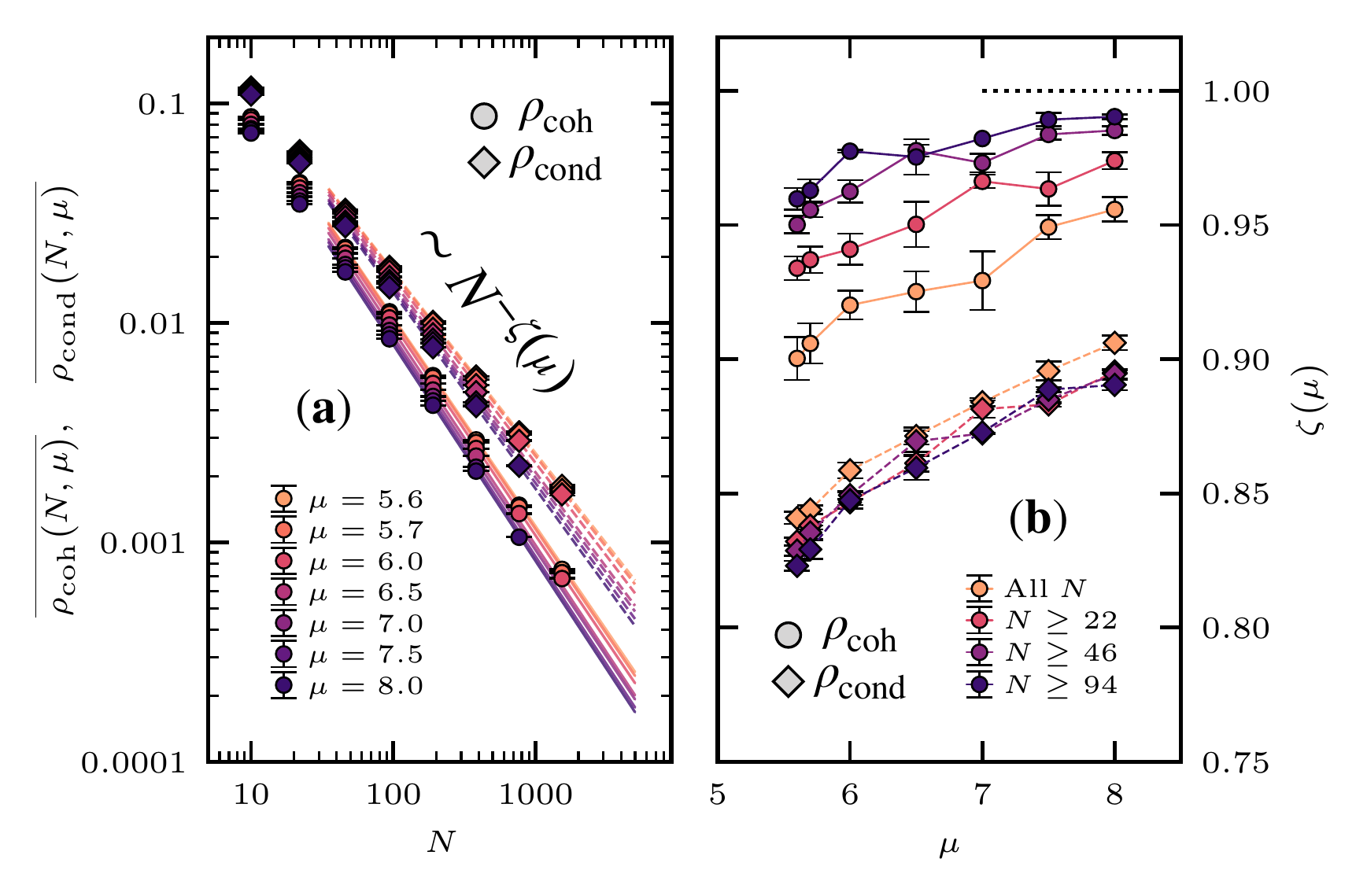}
    \caption{(a) Finite-size decay of the disorder-averaged coherent (circles) and condensate (diamonds) densities in the disordered regime $\mu>5.5$. Lines are power-law fits. (b) Decay exponents $\zeta_\mathrm{coh/cond}$, estimated from fits to the form Eq.~\eqref{eq:zeta}, are plotted against disorder strength $\mu$. Results for different fitting windows show that $\zeta_\mathrm{coh}\to 1$ is expected for large enough system sizes, and $\zeta_\mathrm{cond}(\mu)<1$ for the considered values of disorder strength.}
    \label{fig:zeta}
\end{figure}

In the ordered phase both coherent and BEC densities take finite values of comparable magnitude. However, in the disordered regime, for $\mu> \mu_\mathrm{c}$ the two order parameters vanish in the thermodynamic limit as power-laws
\begin{equation}
    \overline{\rho_\mathrm{coh/cond}}\propto N^{-\zeta_\mathrm{coh/cond}},
    \label{eq:zeta}
\end{equation}
with different decay exponents $\zeta_\mathrm{coh}\neq \zeta_\mathrm{cond}$. This is clear from Fig.~\ref{fig:zeta} where the decay of $\rho_\mathrm{coh}(N)$ is compatible with a conventional $1/N$ behavior while the condensate density shows a slower decay with $\zeta_\mathrm{cond}<1$.

\subsubsection{Phenomenological description for the Bose glass phase}
\label{sec:toy}

\begin{figure}[!t]
    \centering
    \includegraphics[width=0.7\columnwidth]{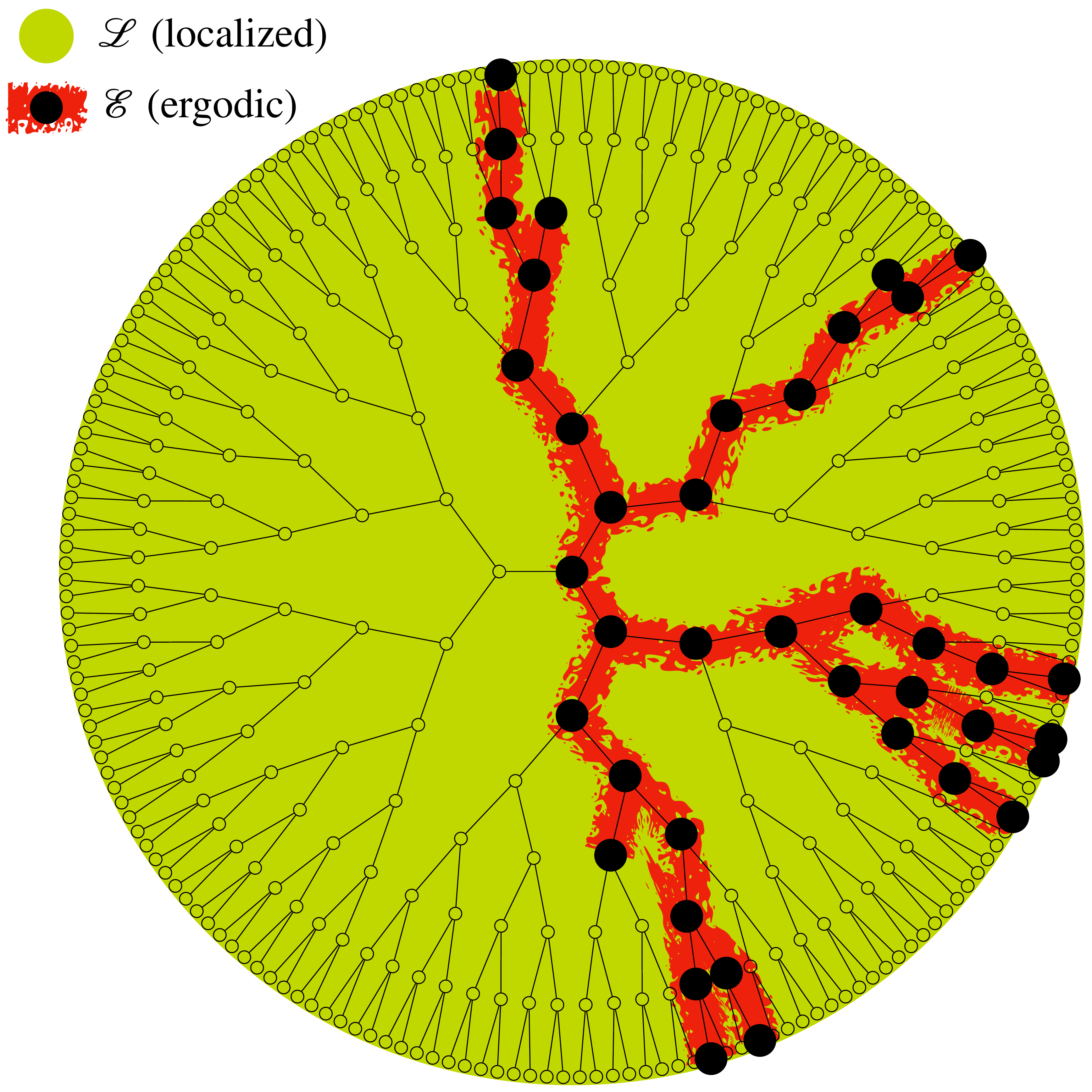}
    \caption{Sketch for the phenomenological description inspired from the QMC results shown in Fig.~\ref{fig:sample_real_space}. The Cayley tree is divided in two subsets: the ergodic part ${\cal{E}}$ where the weights (of both the correlations and the leading orbital) are finite and of comparable magnitude, and the localized subset ${\cal L}$ where instead $\mathsf{C}_{ij}$ and the coefficients $a_i$ are very small, exponentially localized, see Eqs.~\eqref{eq:ai} and \eqref{eq:ansatz_C}.}
    \label{fig:sketch_toy}
\end{figure}

We want to build a phenomenological description which captures all relevant features of the disordered regime. Building on our QMC results, in particular the real-space properties shown in Fig.~\ref{fig:sample_real_space}, and the Refs.~\cite{garciamata2017, PhysRevResearch.2.012020}, we propose a simple two-parameter ansatz which describes both the pair-wise correlations $\mathsf{C}_{ij}$ and the coefficients $a_i$ of the leading orbital $|\phi_1\rangle =\sum_{i=1}^N a_i|i\rangle$ associated to the largest eigenvalue $\lambda_1$. We first model the inhomogeneity, clearly visible in the right panels of Fig.~\ref{fig:sample_real_space}, by dividing the Cayley tree in two subsets, as sketched in Fig.~\ref{fig:sketch_toy}: An ergodic region ${\cal{E}}$ where all the weights are finite and of comparable magnitude, and a localized subset ${\cal L}$ where instead $\mathsf{C}_{ij}$ and the coefficients $a_i$ are very small and decay exponentially.

This behavior is modeled by the following ans\"atze,
\begin{equation}
    |a_i| \sim \left\{
    \begin{array}{rl}
    \exp(-g_{i}/\xi)\quad\quad&\mathrm{if}\quad i\in {\cal{L}}\\
    {\mathrm{constant}}\quad\quad&\mathrm{if}\quad i\in {\cal{E}},
    \end{array}
    \right.
    \label{eq:ai}
\end{equation}
where $g_i$ is the distance across the tree generations from the localization center, and
\begin{equation}
    \mathsf{C}_{ij} \sim \left\{
    \begin{array}{rl}
    \exp(-g_{ij}/\xi)\quad\quad&\mathrm{if}\quad i,j\in {\cal{L}}\\
    {\mathrm{constant}}\quad\quad&\mathrm{if}\quad i,j\in {\cal{E}},
    \end{array}
    \right.
    \label{eq:ansatz_C}
\end{equation}
where $g_{ij}$ is the distance between two sites. If $i$ and $j$ belong to different subsets, we will also assume an exponential decay.
In addition to the length scale $\xi$, which controls the localized part (and which should depend on the disorder strength), we introduce a second disorder-dependent parameter $\alpha$ in order to describe the size of the ergodic support
\begin{equation}
    N_{\cal E}\propto N^{\alpha},
\end{equation}
with $0\le \alpha<1$ in the disordered phase. The size of the localized part $\cal L$ has a dominant scaling $N_{\cal L}\sim N$ (with subleading corrections).

\paragraph{Coherent density---} In order to estimate the  coherent density Eq.~\eqref{eq:rho_coh}, we have to perform a summation over all possible pairs of correlators. Using the ansatz Eq.~\eqref{eq:ansatz_C} we arrive at
\begin{equation}
    \rho_\mathrm{coh}\approx \frac{1}{2N}+\frac{c_1{\mathrm{e}}^{G\left(\ln K-2/\xi\right)}}{N}+\frac{c_2}{N^{2(1-\alpha)}},
\end{equation}
where $c_1$ and $c_2$ are constants. The first term accounts for particle density (half-filling) contribution, the second one comes from the localized support, and the third one from the ergodic part. Using the fact that $G=\ln N / \ln K$, we get a decay exponent governing $\rho_\mathrm{coh}\sim N^{-\zeta_\mathrm{coh}}$
\begin{equation}
    \zeta_\mathrm{coh}=\min\left[1,\frac{2}{\xi\ln K},2(1-\alpha)\right].
    \label{eq:zeta_coh}
\end{equation}
Our QMC results, see Fig.~\ref{fig:zeta}, strongly suggest that $\zeta_\mathrm{coh}=1$, which constraints $\xi\le 2/\ln K$ and $\alpha\le 1/2$.
\paragraph{BEC density---} Then, one can also get an estimate for the largest eigenvalue $\lambda_1$ of the 1BDM
\begin{equation}
    \mathsf{C}|\phi_1\rangle = \lambda_1|\phi_1\rangle.
\end{equation}
We have to solve
\begin{equation}
    a_0\mathsf{C}_{00}+a_1\mathsf{C}_{01}+\cdots+a_{N-1}\mathsf{C}_{0N-1}=\lambda_1a_0,
\end{equation}
which, using Eqs.~\eqref{eq:ai} and~\eqref{eq:ansatz_C}, the fact that $\xi<2/\ln K$, and after a proper normalization of the leading orbital (see below), yields for the dominant term
\begin{equation}
    \lambda_1\propto N^{\alpha/2}.
\end{equation}
This gives a BEC density $\rho_\mathrm{cond}=\lambda_1/N\propto N^{-1+\alpha/2}$, and therefore a decay exponent
\begin{equation}
    \zeta_\mathrm{cond}=1-\frac{\alpha}{2}
    \label{eq:zeta_cond}
\end{equation}
\paragraph{Participation entropies---} The third quantity which can be estimated from our phenomenological description is the participation entropy of the leading orbital, as previously  defined in Eq.~\eqref{eq:part_ent_sq}. Using the fact that $\xi<2/\ln K$, the normalization of the ansatz wave-function Eq.~\eqref{eq:ai} yields
\begin{equation}
    |a_i| \propto N^{-\alpha/2}\times\left\{
    \begin{array}{rl}
    \exp(-g_{i}/\xi)\quad\quad&\mathrm{if}\quad i\in {\cal{L}}\\
    {\mathrm{constant}}\quad\quad&\mathrm{if}\quad i\in {\cal{E}},
    \end{array}
    \right.
\end{equation}
The $q$-R\'enyi entropies will depend on a threshold index
\begin{equation}
    q^*=\frac{\xi\ln K}{2}(1-\alpha)<1,
    \label{eq:qstar}
\end{equation}
such that
\begin{equation}
    S_q\approx\begin{cases}
        \frac{1-q\left(\alpha+\frac{2}{\xi\ln K}\right)}{1-q}\ln N &{\mathrm{if~}}\quad q<q^*\\
        \alpha\ln N &{\mathrm{if~}}\quad q>q^*,
    \end{cases}
    \label{eq:Sqi}
\end{equation}

\begin{table*}[!t]
    \vspace*{5mm}
    \begin{minipage}{1.8\columnwidth}
        \center
        \begin{ruledtabular}
            \begin{tabular}{lccc}
                \thead{} & \thead{Coherent density\\ $\rho_\mathrm{coh}\propto N^{-\zeta_\mathrm{coh}}$} & \thead{Condensed density\\ $\rho_\mathrm{cond}\propto N^{-\zeta_\mathrm{cond}}$} & \thead{Participation entropies\\ $S_q\approx D_q\ln N$}
                \\\hline\\[-0.5em]
                \makecell{QMC} & \makecell{$\zeta_\mathrm{coh}\approx 1$ (Fig.~\ref{fig:zeta})} & \makecell{$\zeta_\mathrm{cond}< 1$ (Fig.~\ref{fig:zeta})} & \makecell{$0<D_q<1$ (Fig.~\ref{fig:ent_cond_mode})}\\[-0.5em]
                \\\hline\\[-0.8em]
                \makecell{Phenomenological\\ description}& \makecell{$\zeta_\mathrm{coh}=1$\\ if~$\xi<\frac{2}{\ln K}$ and $\alpha<1/2$} & \makecell{$\zeta_\mathrm{cond}=1-\frac{\alpha}{2}$\\ if~$\xi<\frac{2}{\ln K}$} & \makecell{$D_q\approx\begin{cases}
                    \frac{1-q\left(\alpha+\frac{2}{\xi\ln K}\right)}{1-q} &{\mathrm{if~}}\quad q<q^*\\
                    \alpha &{\mathrm{if~}}\quad q>q^*,
                \end{cases}$}\\
            \end{tabular}
        \end{ruledtabular}
    \end{minipage}
    \caption{Summary of some properties of the disordered phase of dirty bosons on a Cayley tree with branching number $K$. QMC estimates are shown, together with analytical results obtained from a two-parameter phenomenological description (see Sec.~\ref{sec:toy}). The decay exponents $\zeta_{\mathrm{coh/cond}}$ of both coherent and condensate densities are displayed, together with the (multi)fractal dimension $D_q$ governing the ergodicity properties of the leading orbital of the 1BDM. The threshold R\'enyi index $q^*$ is given in Eq.~\eqref{eq:qstar}.}
    \label{tab:toy}
\end{table*}

\paragraph{Consequences and comparison with QMC---} From QMC simulations, see Fig.~\ref{fig:zeta}, we expect for the disordered regime that $\zeta_\mathrm{coh}=1$ and $\zeta_\mathrm{cond}<1$, a behavior perfectly reproduced by our phenomenological description, see Eqs.~\eqref{eq:zeta_coh} and~\eqref{eq:zeta_cond}, provided that the characteristic length scale $\xi<2/\ln K$, and the parameter $\alpha<1/2$.  Moreover, the leading orbital $|\phi_1\rangle$ was found to be delocalized and non-ergodic (see Fig.~\ref{fig:ent_cond_mode}) with a multifractal behavior at small $q$, followed by a simpler fractal behavior at larger $q$ with a constant $D_q<1$. Here, our phenomenological description is also able to capture such a behavior, see Eq.~\eqref{eq:Sqi}, with a threshold R\'enyi parameter $q^*<1$, in agreement with QMC results. The physical interpretation of such (multi) fractal properties is fairly simple: large values of $q$ probe the larger components of $|\phi_1\rangle$, thus essentially exploring the ergodic subset $\cal E$ (see Fig.~\ref{fig:sketch_toy}) of size $N^\alpha$. Conversely, at small $q$, the localized component cannot be ignored, and will also contributes to the multifractal dimension. In Tab.~\ref{tab:toy} we give a summary of these findings, and a comparison betwen the phenomenological description and QMC results.

It is also worth mentioning that a simple geometric interpretation of the exponent $\alpha$ can be done, in terms of an effective branching number $K_\mathrm{eff}$. Indeed, instead of having all branches of the tree equally contributing, we allow a reduced branching parameter $K_\mathrm{eff}=pK$ with $1/K<p\le1$~\footnote{The restriction $p>1/K$ ensures the existence of a percolating cluster.}. The parameter $p$ can be interpreted as the probability to follow a branch. Then we have the simple relation $\alpha=1+\ln p/\ln K$.

Finally, this phenomenological description provides a nice physical picture for the anomalous power-law scaling $\lambda_1\sim N^{1-\zeta_\mathrm{cond}}$ (with $\zeta_\mathrm{cond}<1$) which is a direct consequence of the fractal behavior of the associated orbital. At large enough $q$, where all fractal dimensions are the same $D_q\equiv D$, we have the very simple result
\begin{equation}
    1-\zeta_\mathrm{cond}={D/2}.
    \label{eq:D2}
\end{equation}
A direct comparison of the phenomenological description results with QMC simulations gives a surprisingly accurate agreement with Eq.~\eqref{eq:D2}. For instance, at the strongest disordered strength that we have considered ($\mu=8$), the QMC estimate for $1-\zeta_\mathrm{cond}=0.10(1)$ perfectly agrees with $D_2/2=0.10(1)$. Even more interestingly, the agreement remains quantitative up to criticality (where the quantitative character of the phenomenological description could become questionable), with a QMC estimate at $\mu_\mathrm{c}=5.5$ of $1-\zeta_\mathrm{c}^\mathrm{cond}= 0.20(5)$ which nicely compares with $D_2^\mathrm{c}/2= 0.19(3)$. This observation strongly suggests that the critical point itself is non-ergodic.

At very strong disorder, where QMC simulations become practically impossible to perform because of prohibitively long simulation times, we expect a fully localized phase with $D\to 0$ and $\zeta_\mathrm{cond}\to 1$. However, it is not clear whether or not a second transition towards such a fully localized phase will take place at a larger but finite disorder strength, or perhaps more likely only in the limit of infinite randomness.

\section{Discussions and conclusion}
\label{sec:conclusion}

\subsection{Summary of our results}

In this paper, building on large-scale quantum Monte Carlo simulations, we have investigated the zero-temperature phase diagram of hard-core bosons in a random potential on site-centered Cayley trees with branching number $K=2$. We find that the system undergoes a disorder-induced quantum phase transition at finite disorder strength $\mu_\mathrm{c}\approx 5.5(5)$ between a long-range ordered Bose-Einstein condensed state at weak disorder and a disordered Bose glass regime at stronger disorder.

We characterize the two different phases and the critical properties of the transition using several physical quantities: The gap ratio from the largest occupation numbers of the one-body density matrix, the local densities of bosons, the off-diagonal correlations, the coherent density $\rho_\mathrm{coh}$, the condensed density $\rho_\mathrm{cond}$ as well as the leading orbital and its participation entropy. We have performed a careful scaling analysis on the last three quantities as the transition is approached from the weak disorder side $\mu<\mu_\mathrm{c}$. In the strong disorder side $\mu>\mu_\mathrm{c}$, we have described the physics using a phenomenological description.

All the observations and analyzes agree on the same physical image of the transition. At low disorder, there is a characteristic volume $\Lambda$ beyond which the system shows off-diagonal long-range order: The long-distance correlator  is constant ; its distribution is stationary ; the coherent density  is finite $\rho_\mathrm{coh}>0$ as well as the  condensed density $\rho_\mathrm{cond}>0$ ; the leading orbital (i.e., the condensed mode) is ergodic with a multifractal dimension $D_2=1$. In the strong disorder regime, we observe clear signatures of a non-ergodic Bose glass phase: the typical and average correlators decrease exponentially with different localization lengths, their distribution has a traveling wave regime and a large power-law tail $P(\mathsf{C})\sim\mathsf{C}^{-(1+B)}$ with an exponent $B<1$, a signature of replica symmetry breaking, a crucial glassy property. The coherent density decreases as $\rho_\mathrm{coh}\sim1/N$ while the condensed density decreases with a non-trivial power-law $\rho_\mathrm{cond}\sim N^{-\zeta_\mathrm{cond}}$. Moreover, the leading orbital of the one-body density matrix is multifractal ($D_q<1$).

These observations can be precisely accounted for by a simple phenomenological description for the disordered regime where the Cayley tree is divided in two subsets: A pruned tree (with $N^\alpha$ sites) along which the correlator shows long-range order and the leading orbital is delocalized, while the remaining sites show a strong exponential localization. In our picture, the Bose glass phase is similar to the non-ergodic delocalized phase of the Anderson transition on the Cayley tree, where similar multifractal properties have been predicted in a broad range of disorder~\cite{monthus2011,tikhonov2016,sonner2017,parisi2019, kravtsov2018,biroli2018,facoetti2016,savitz2019}. The two characteristic scales $\xi_\mathrm{typ}$ and $\alpha$ characterizing the Bose glass phase are also reminiscent of the two localization lengths that govern the localized phase of the Anderson transition on random graphs \cite{PhysRevResearch.2.012020}. In particular, different observables are governed by different characteristic scales. While $\rho_\mathrm{coh}$  is controlled by the bulk localization properties, i.e., $\xi_\mathrm{typ}$, $\rho_\mathrm{cond}$ and the leading orbital are dominated by the rare delocalized pruned tree, thus by the scale $\alpha$.

The comparison of our results with the predictions of the cavity mean field~\cite{feigelman2010} clearly indicates a new condensed ergodic phase at low disorder which is absent in the cavity approach. It remains to be studied if the Bose glass phase that we observe, which is a non-ergodic delocalized phase, corresponds to that predicted by cavity approach and if a second transition to a completely localized phase occurs at stronger disorder.
Another possibility is that the cavity mean field describes different physics, because of the approximation made when dealing with the Ising model, which is clearly different from our $\mathrm{U}(1)$ symmetric bosonic system.

Finally, the non-trivial scaling laws that we found suggest that there is no finite upper critical dimension $d_\mathrm{c}$ beyond which conventional onset of mean-field theory would take place.

\subsection{Open questions}

This work is the first of its kind, studying by an unbiased numerical method the dirty boson problem on an effectively infinite-dimensional lattice. While we address several fundamental points such as the existence of a quantum phase transition at finite disorder strength and the nature of some of its critical properties, several questions remain open.

A first one concerns the critical properties of the transition when approached from the localized phase $\mu>\mu_\mathrm{c}$. For instance, different scaling laws on both sides of the transition were found for the Anderson localization transition on random graphs (which are also effectively infinite dimensional)~\cite{garciamata2017}: a volumic scaling on the delocalized side and a scaling with the linear size of the system on the localized side. Would the same phenomenology apply for boson localization? However, accessing strong disorder with quantum Monte Carlo is computationally challenging and expensive, which is why we limited our scaling analysis to the ordered phase.

Another interesting question concerns the possible universal properties of the delocalization-localization transiton in infinite dimension. The Cayley tree that we studied in this paper is one example of such effectively infinite dimensional lattice, but other graphs meet the requirements, such as random graphs or small-world networks. In particular, by studying the dirty boson problem on one of these lattices would allow to quantify the effect of geometrical loops, absent on the Cayley tree, and to quantify the effect of the extensive number of boundary lattice sites, specific to the Cayley tree.

A peculiar property of the superfluid--Bose glass transition is the predicted~\cite{fisher1989} hyperscaling relation $z=d$, between the dimension of the system $d$, and the dynamical exponent $z$, numerically verified in for $d\le 3$~\cite{alvarez2015,ng2015,hitchcock2006,yu2012b,yao2014,dupont2017}. The exponent $z$ is not readily available from the quantities we considered in this work, as it is usually inferred from the scaling of the superfluid density or the imaginary-time off-diagonal correlation function. Yet, accessing it would be interesting in order to complete the critical properties description of the transition in infinite dimension, and check on the validity of the hyperscaling relation $z=d$ in infinite dimension, in the absence of a finite upper critical dimension.

Regarding the critical exponent $\zeta$, we have identified the hyperscaling relation $\zeta=2\beta/\nu$, which in principle is valid for both coherent and BEC densities. However the fact that at criticality we observed $\zeta_\mathrm{coh}^\mathrm{c}\neq \zeta_\mathrm{cond}^\mathrm{c}$ may imply two different order parameter exponents $\beta_\mathrm{coh}\neq\beta_\mathrm{cond}$.

Finally, our results could suggest an avalanche scenario for the delocalization transition when $\mu\to\mu_\mathrm{c}$ is approached from the Bose glass regime, a process shown rigorously for the Anderson transition on the Cayley tree \cite{aizenman2011,aizenman2011b} and crucially important in the many-body localization transition~\cite{thiery2018}. In our case, it may occur when the exponential bulk localization  does not compensate the exponential increase of the number of sites with the distance, i.e. when $\xi_\mathrm{typ}>\xi_\mathrm{typ}^\mathrm{c}$ with $\xi_\mathrm{typ}^\mathrm{c}$ a critical value which depends only on the branching number.

\begin{acknowledgments}
M.D. was supported by the U.S. Department of Energy, Office of Science, Office of Basic Energy Sciences, Materials Sciences and Engineering Division under Contract No. DE-AC02-05-CH11231 through the Scientific Discovery through Advanced Computing (SciDAC) program (KC23DAC Topological and Correlated Matter via Tensor Networks and Quantum Monte Carlo). N.L. and G.L are supported by the French National Research Agency (ANR) under projects THERMOLOC ANR-16-CE30-0023-02, MANYLOK ANR-18-CE30-0017 and GLADYS ANR-19-CE30-0013, and the EUR grant NanoX No. ANR-17-EURE-0009 in the framework of the ``Programme des Investissements d'Avenir''. This research used the Lawrencium computational cluster resource provided by the IT Division at the Lawrence Berkeley National Laboratory (supported by the Director, Office of Science, Office of Basic Energy Sciences, of the U.S. Department of Energy under Contract No. DE-AC02-05CH11231). This research also used resources of the National Energy Research Scientific Computing Center (NERSC), a U.S. Department of Energy Office of Science User Facility operated under Contract No. DE-AC02-05CH11231. Additionally, the numerical calculations benefited from the high-performance computing resources provided by CALMIP (Grants No. 2018- P0677 and No. 2019-P0677) and GENCI (Grant No. 2018- A0030500225).
\end{acknowledgments}

\appendix

\section{Additional information on the quantum Monte Carlo simulations}
\label{app:qmc_conv}

The quantum Monte Carlo data displayed in the main text is computed at the inverse temperature $\beta=1/T$ reported in Tab.~\ref{tab:qmc_parameters}. The number of disorder samples $N_\mathrm{s}$ for each system size is also reported.

\begin{table}[!h]
    \vspace*{5mm}
    \begin{minipage}{0.6\columnwidth}
        \center
        \begin{ruledtabular}
            \begin{tabular}{cccc}
                \thead{$G$} & \thead{$N$} & \thead{$\beta$} & \thead{$N_\mathrm{s}$}\\
                \hline
                \makecell{$2$} & \makecell{$10$} & \makecell{$32$} & \makecell{$>2000$}\\
                \makecell{$3$} & \makecell{$22$} & \makecell{$32$} & \makecell{$>2000$}\\
                \makecell{$4$} & \makecell{$46$} & \makecell{$64$} & \makecell{$>2000$}\\
                \makecell{$5$} & \makecell{$94$} & \makecell{$64$} & \makecell{$>2000$}\\
                \makecell{$6$} & \makecell{$190$} & \makecell{$128$} & \makecell{$>2000$}\\
                \makecell{$7$} & \makecell{$382$} & \makecell{$128$} & \makecell{$>2000$}\\
                \makecell{$8$} & \makecell{$766$} & \makecell{$256$} & \makecell{$>1000$}\\
                \makecell{$9$} & \makecell{$1534$} & \makecell{$256$} & \makecell{$>500$}\\
                \makecell{$10$} & \makecell{$3070$} & \makecell{$512$} & \makecell{$>300$}\\
            \end{tabular}
        \end{ruledtabular}
    \end{minipage}
    \caption{Inverse temperature $\beta=1/T$ used in the quantum Monte Carlo algorithm depending on the system size (number of generations $G$, total number of lattice sites $N$). The number of independent disordered samples $N_\mathrm{s}$ computed to perform the disorder average is also reported.}
    \label{tab:qmc_parameters}
\end{table}

\subsection{Convergence with temperature}

\begin{figure}[!t]
    \centering
    \includegraphics[width=1.0\columnwidth]{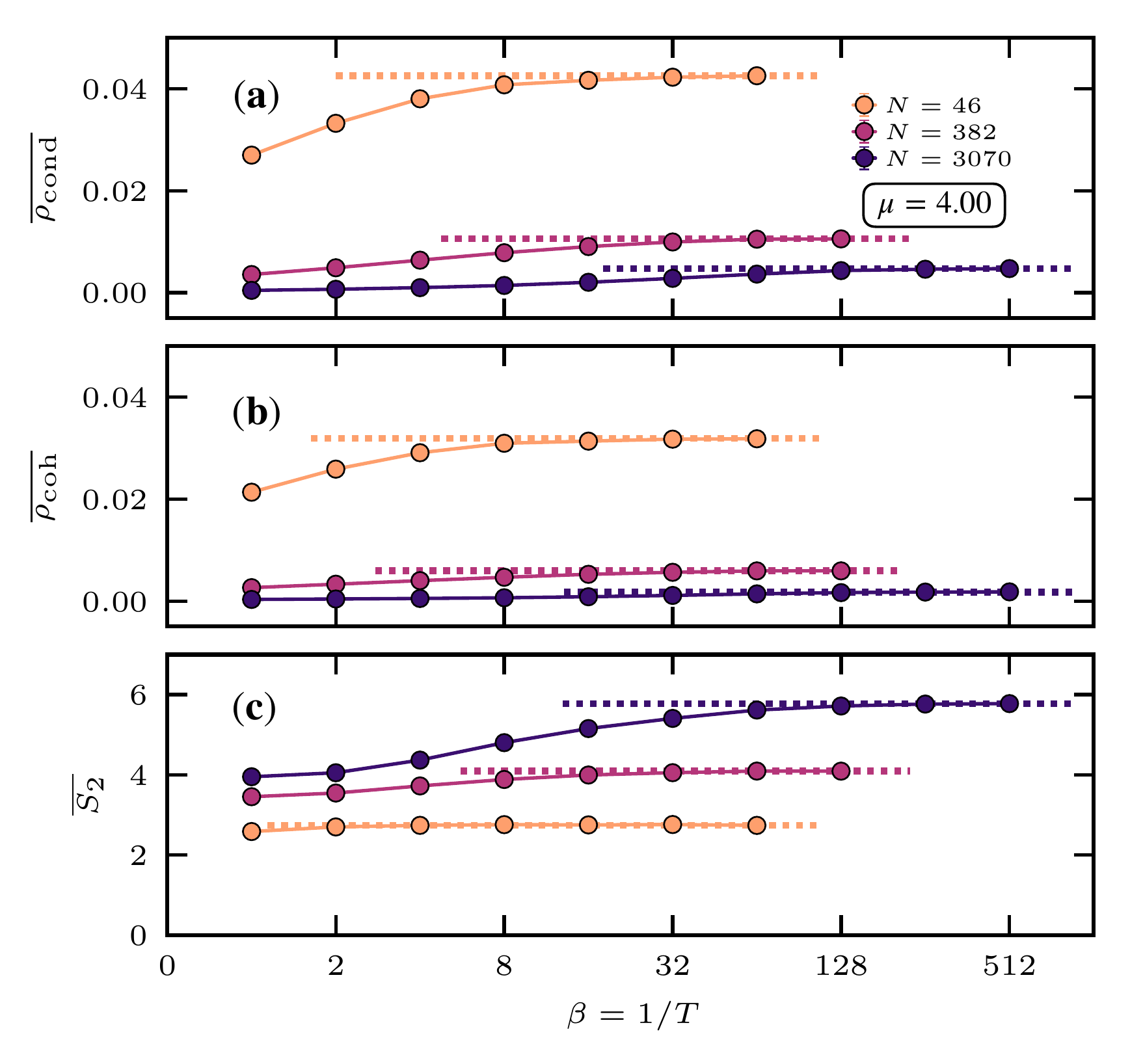}
    \caption{Convergence of the disorder-averaged condensed density $\overline{\rho_\mathrm{cond}}$, the disorder-averaged coherent density $\overline{\rho_\mathrm{coh}}$, and the disorder-averaged participation entropy $\overline{S_2}$ versus the inverse temperature $\beta=1/T$. Three system sizes are displayed $N=46$ ($N_\mathrm{s}=2000$), $N=382$ ($N_\mathrm{s}=2000$) and $N=3070$ ($N_\mathrm{s}=360$) at disorder strength $\mu=4$. Note the logarithmic scale fo the $x$-axis. The data displayed in the main text are those at the largest inverse temperature, see Tab.~\ref{tab:qmc_parameters}.}
    \label{fig:conv_vs_beta}
\end{figure}

The stochastic series expansion quantum Monte Carlo is a finite temperature method. Therefore, it is important to perform calculations at sufficiently low temperatures to capture ground state properties. For three representative system sizes $N=46$, $N=382$ and $N=3070$ at disorder strength $\mu=4$, we show in Fig.~\ref{fig:conv_vs_beta} the convergence of the disorder-averaged condensed density $\overline{\rho}$ and the disorder-averaged participation entropy $\overline{S_2}$ versus the inverse temperature $\beta=1/T$. The data displayed in the main text are those at the largest inverse temperature, as indicated in Tab.~\ref{tab:qmc_parameters}. We have found that these temperatures are low enough to reliably probe the ground state in the quantum Monte Carlo simulations.

\subsection{Convergence with number of samples}

\begin{figure}[!t]
    \centering
    \includegraphics[width=1.0\columnwidth]{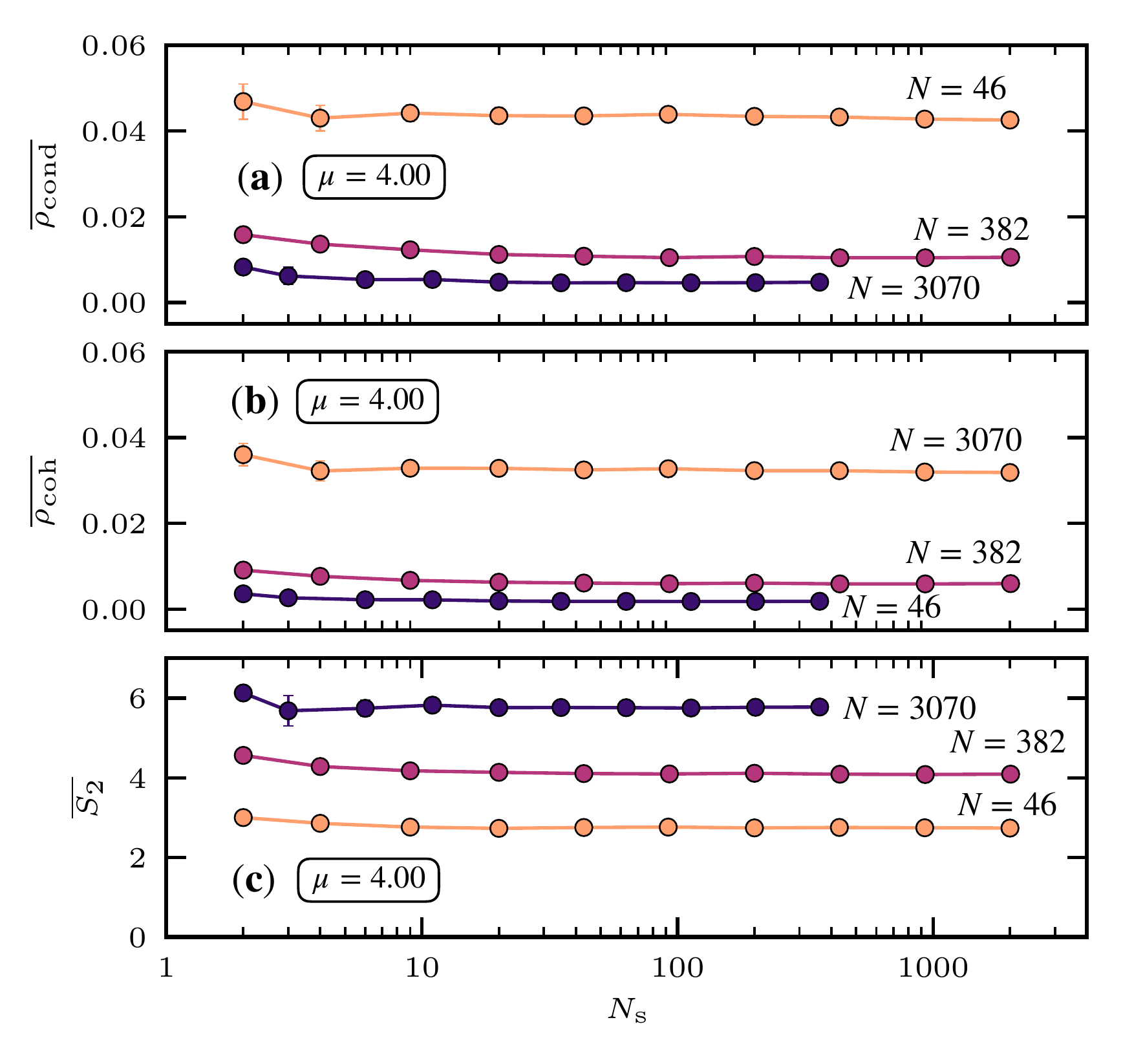}
    \caption{Convergence of the disorder-averaged condensed density $\overline{\rho_\mathrm{cond}}$, the disorder-averaged coherent density $\overline{\rho_\mathrm{coh}}$, and the disorder-averaged participation entropy $\overline{S_2}$ versus the number of independent samples $N_\mathrm{s}$ considered to perform the average. Three system sizes are displayed $N=46$, $N=382$ and $N=3070$ at disorder strength $\mu=4$. See Tab.~\ref{tab:qmc_parameters}.}
    \label{fig:conv_vs_samples}
\end{figure}

The convergence with the number of disorder samples is checked by performing averages including an increasing number of samples $N_\mathrm{s}$. We show in Fig.~\ref{fig:conv_vs_samples} that convergence is quickly achieved (with a few tens of samples) by considering three representative system sizes are displayed $N=46$, $N=382$ and $N=3070$ at disorder strength $\mu=4$. Even at stronger disorder, a few hundred of samples is sufficient to obtain reliable average estimates. We consider in general $>2000$ samples, except for the largest system sizes due to the numerical cost of simulating them, see Tab.~\ref{tab:qmc_parameters}.

\section{Finite-size scaling analysis}
\label{app:scaling}

\begin{figure}[!t]
    \centering
    \includegraphics[width=0.8\columnwidth]{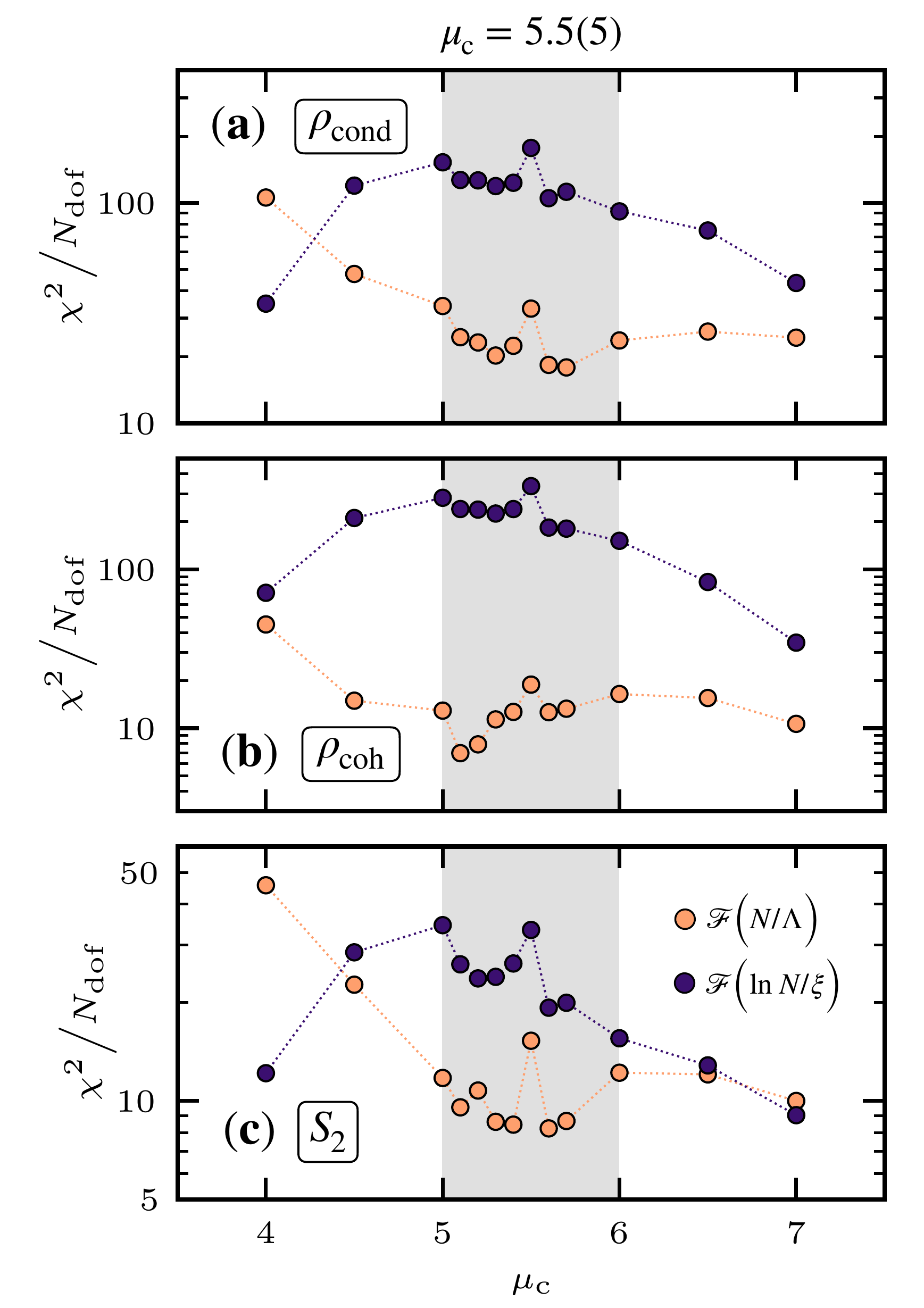}
    \caption{Chi-squared statistic $\chi^2$ per degree of freedom $N_\mathrm{dof}$ for the best volumic and linear fits of the data ($\mu<\mu_\mathrm{c}$) obtained following a scaling of the form of Eq.~\eqref{eq:scafunc} versus the critical disorder strength $\mu_\mathrm{c}$ for (a) the condensed density, (b) the coherent density and (c) the participation entropy of the leading eigenmode. The volumic scaling is systematically better, with a minimum around $\mu_\mathrm{c}\approx 5.5\pm 0.5$. Refer to discussion of Sec.~\ref{sec:quantum_critical} in the main text.}
    \label{fig:chi2_collapse}
\end{figure}

To determine the value of the critical disorder strength $\mu_\mathrm{c}$ and the critical properties of the transition for $\mu<\mu_\mathrm{c}$, we perform a finite-size scaling analysis, as detailed in Sec.~\ref{sec:quantum_critical}. For the condensed density, the coherent density and the participation entropy of the leading orbital, both a linear and volumic scaling hypothesis are tested. Their quality is measured by the chi-squared statistic $\chi^2$ per degree of freedom $N_\mathrm{dof}$ of fitting the numerical data to the corresponding scaling function, which also depends on the choice of critical disorder strength $\mu_\mathrm{c}$ considered. Hence, by plotting $\chi^2/N_\mathrm{dof}$ versus $\mu_\mathrm{c}$, we are able to estimate the best scaling hypothesis and locate the transition. As shown in Fig.~\ref{fig:chi2_collapse}, the volumic scaling is systematically better compared to the linear one, with a minimum observed for all quantities around $\mu_\mathrm{c}\approx 5.5\pm 0.5$.

\bibliography{bibliography,refs_SW}

\end{document}